\documentclass[preprint,12pt]{elsarticle}

\usepackage{amsmath}
\usepackage{amssymb}

\usepackage{siunitx}
\usepackage{svrsymbols}

\usepackage{hyperref}

\usepackage{tikz}
\usetikzlibrary{3d,perspective}

\usepackage{placeins}

\usepackage{caption}
\usepackage{subcaption}

\journal{Nuclear Instruments and Methods in Physics Research A, \space}

\begin{document}

\begin{frontmatter}

    
    
    \title{Performance study of the JadePix-3 telescope from a beam test}
    
    
    \author[ihep]{Sheng Dong}
    \author[ustc1,ustc2]{Zhiliang Chen}
    \author[ucas,ihep]{Jia Zhou\corref{cor2}}
    \cortext[cor2]{Graduated in the summer 2023}
    \author[jlu]{Xingye Zhai}
    \author[ucas]{Anqi Wang}
    \author[ucas]{Yunxiang Wang}
    \author[ccnu]{Hulin Wang}
    \author[ustc1,ustc2]{Lailin Xu}
    \author[ihep]{Jing Dong}
    \author[ihep]{Yang Zhou}
    \author[ihep]{Yunpeng Lu\corref{cor1}}
    \cortext[cor1]{Corresponding author}
    \ead{yplu@ihep.ac.cn}
    \author[ihep,ucas]{Mingyi Dong}
    \author[ihep,ucas]{Hongyu Zhang}
    \author[ihep,ucas]{Qun Ouyang\corref{cor1}}
    \ead{ouyq@ihep.ac.cn}
    
    \affiliation[ihep]{organization={State Key Laboratory of Particle Detection and Electronics, Institute of High Energy Physics, Chinese Academy of Sciences},
        city={BeiJing},
        postcode={100049},
        state={BeiJing},
        country={China}}
    
    \affiliation[ustc1]{organization={State Key Laboratory of Particle Detection and Electronics, University of Science and Technology of China},
        city={Hefei},
        postcode={230026},
        state={Anhui},
        country={China}}
    
    \affiliation[ustc2]{organization={Department of Modern Physics, University of Science and Technology of China},
        city={Hefei},
        postcode={230026},
        state={Anhui},
        country={China}}
    
    \affiliation[ucas]{organization={University of Chinese Academy of Sciences},
        city={Beijing},
        postcode={100049},
        state={Beijing},
        country={China}}
    
    \affiliation[jlu]{organization={Jilin University},
        city={Changchun},
        postcode={130012},
        state={Jilin},
        country={China}}
    
    \affiliation[ccnu]{organization={PLAC, Key Laboratory of Quark \& Lepton Physics (MOE), Central China Normal University},
        city={Wuhan},
        postcode={430079},
        state={Hubei},
        country={China}}
    
    \begin{abstract}
        We present the results of a beam test conducted on a telescope using the JadePix-3 pixel sensor, developed with TowerJazz 180 nm CMOS imaging technology. The telescope is composed of five planes, each equipped with a JadePix-3 sensor with pitches of $26\times16$ \si{\um^2} and $23.11\times16$~\si{\um^2}. In addition, it features an FPGA-based synchronous readout system. The telescope underwent testing using an electron beam with energy ranging from 4 to \SI{6}{GeV}. At an electron energy of \SI{5.4}{GeV}, the telescope demonstrated superior spatial resolutions of 2.6 and \SI{2.3}{\um} in two dimensions. By designating the central plane as the device under test, we evaluated the JadePix-3 sensor's spatial resolutions as 5.2 and \SI{4.6}{\um} in two dimensions, achieving a detection efficiency of over \SI{99.0}{\percent}.
    \end{abstract}
    
    
    
    
\end{frontmatter}



\FloatBarrier
\section{Introduction}
\label{intro}
The Circular Electron Positron Collider (CEPC) is designed to investigate the Higgs boson and conduct essential tests of the fundamental principles underlying the Standard Model, potentially revealing new physics. As a prospective experimental configuration, it imposes stringent technical demands on a variety of detectors. Specifically, the vertex detector necessitates that a single-point resolution of the first layer must be better than \SI{3}{\um}, and the power consumption of the sensors and readout electronics should not exceed \SI{50}{\mW/{cm}^2}\cite{cepc2018cepcVol1, cepc2018cepcVol2}.

\begin{figure}[h]
    \centering
    \begin{subfigure}[t]{0.57\textwidth}
        \begin{tikzpicture}[scale=0.65]
            \draw[fill=lightgray] (0, 0) rectangle (6.1, 10.4);
            \draw[] ( 1.05, 1.51) rectangle (5.9, 9.7);
            \draw[draw=none,fill=cyan] ( 1.05, 1.51) rectangle (1.05+2*4.85/4, 9.7);
            \draw[draw=none,fill=teal] ( 1.05+2*4.85/4, 1.51) rectangle (1.05+3*4.85/4, 9.7);
            \draw[draw=none,fill=cyan] ( 1.05+3*4.85/4, 1.51) rectangle (1.05+4*4.85/4, 9.7);
            
            \draw[dotted] (1.05+4.85/4, 1.51) -- (1.05+4.85/4, 9.7);
            \draw[dotted] (1.05+2*4.85/4, 1.51) -- (1.05+2*4.85/4, 9.7);
            \draw[dotted] (1.05+3*4.85/4, 1.51) -- (1.05+3*4.85/4, 9.7);
            
            \draw[dotted] (0,0) -- (0, -0.5);
            \draw[dotted] (1.05,0) -- (1.05, -0.5);
            
            \draw[dotted] (-0.5,0) -- (0, 0);
            \draw[dotted] (0.7,1.51) -- (1.05, 1.51);
            \node[] at (0.5, -0.3) {\tiny 1.05};
            \draw[to-] (0, -0.3)--(0.15, -0.3);
            \draw[-to] (0.85, -0.3)--(1.05, -0.3);
            
            \node[] at (0.5, -0.3) {\tiny 1.05};
            \node[] at (3.5, -0.3) {\tiny 5.05};
            \draw[to-] (1.05, -0.3)--(3.1, -0.3);
            \draw[-to] (3.9, -0.3)--(6.1, -0.3);
            
            \node[rotate=90] at (-0.4, 0.7) {\tiny 1.51};
            \node[rotate=90] at (-0.4, 5.5) {\tiny 8.89};
            
            \draw[to-] (-0.4, 1.51)--(-0.4, 5.0);
            \draw[-to] (-0.4, 5.9)--(-0.4, 10.4);
            
            \draw[to-] (-0.4, 0)--(-0.4, 0.3);
            \draw[-to] (-0.4, 1.0)--(-0.4, 1.51);
            
            \draw[dotted] (0.7,9.7) -- (1.05, 9.7);
            \draw[dotted] (-0.5,10.4) -- (1.05, 10.4);
            
            \draw[dotted] (1.05, 1.51) -- (1.05, 1.21);
            \draw[dotted] (5.9, 1.51) -- (5.9, 1.21);
            \node[] at (3.5, 1.3) {\tiny 4.85};
            \draw[to-] (1.05, 1.31)--(3.1, 1.31);
            \draw[-to] (3.9, 1.31)--(5.9, 1.31);
            
            \draw[dotted] (6.1, 0) -- (6.1, -0.5);
            
            \node[rotate=90] at (0.7, 5.4) {\tiny 8.19};
            \draw[to-] (0.7, 1.51)--(0.7, 5.0);
            \draw[-to] (0.7, 5.8)--(0.7, 9.7);
            
            \node[rotate=90] at (1.6, 5.4) {Sector 0};
            \node[rotate=90] at (1.6+1.2125, 5.4) {Sector 1};
            \node[rotate=90] at (1.6+2*1.2125, 5.4) {Sector 2};
            \node[rotate=90] at (1.6+3*1.2125, 5.4) {Sector 3};
            
            \draw[thick,->,red] (-1, 11) -- (6.8, 11)  node[anchor=south] {Column};
            
            \draw[thick,->,red] (-1, 11) -- (-1, -0.7)  node[anchor= east] {Row};
            
            \node[red] at (-1.5, 11.5) {(0, 0)};
        \end{tikzpicture}
        \caption{}
    \end{subfigure}
    \hfill
    \begin{subfigure}[t]{0.42\textwidth}
        \includegraphics[width=\textwidth]{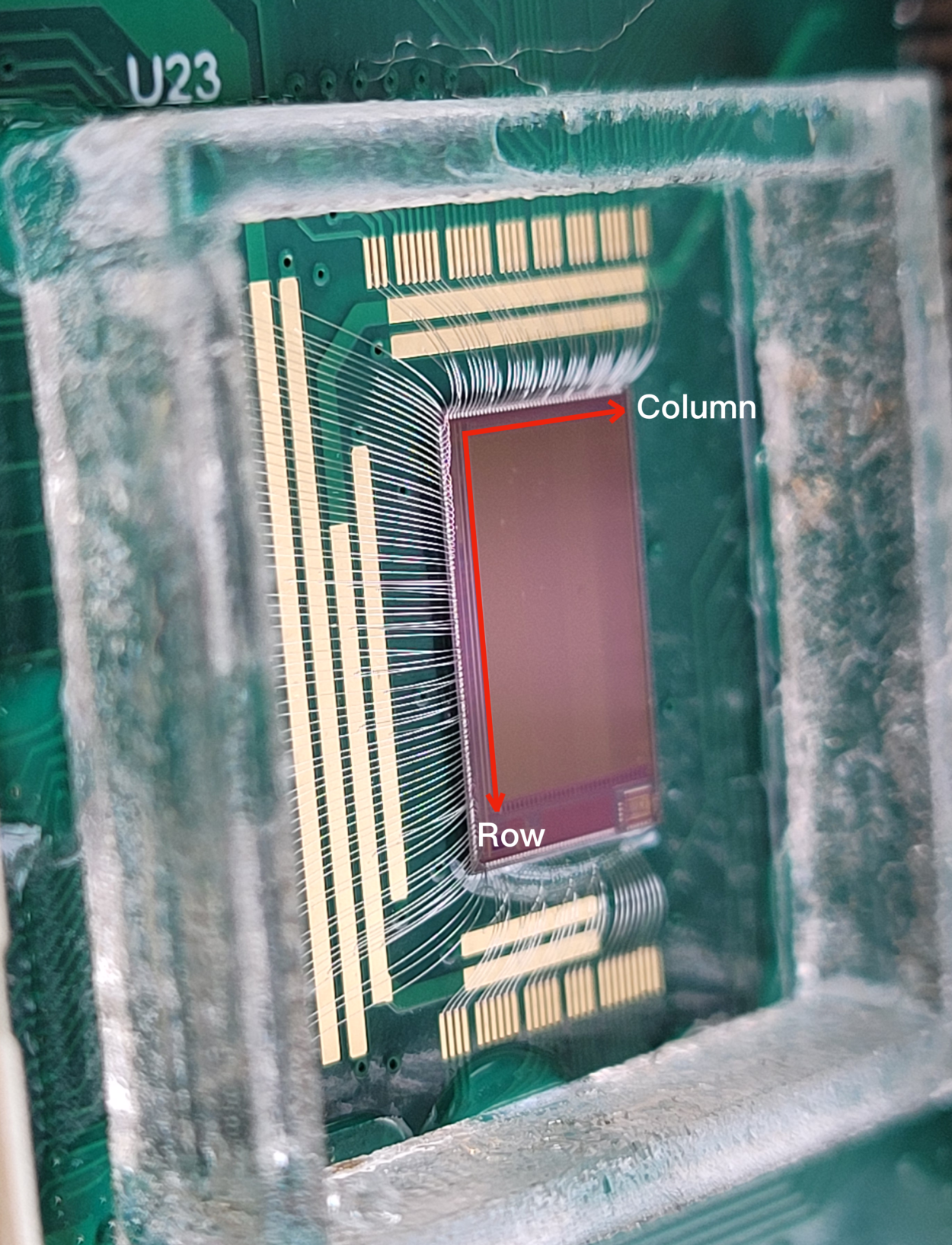}
        \caption{}
    \end{subfigure}
    \caption{(a) JadePix-3 layout. The gray area illustrates the pixel periphery, whereas the cyan and teal areas represent the pixel array. Sectors 0, 1, and 3 feature a pitch of $26\times16.0$ \si{\um^2}, whereas sector 2 has a pixel pitch of $23.11\times16.0$ \si{\um^2}. Each sector's pixel matrix comprises 48 columns and 512 rows. (b) Photograph of a wire-bonded chip.}
    \label{fig:jadepix3-layout}
\end{figure}
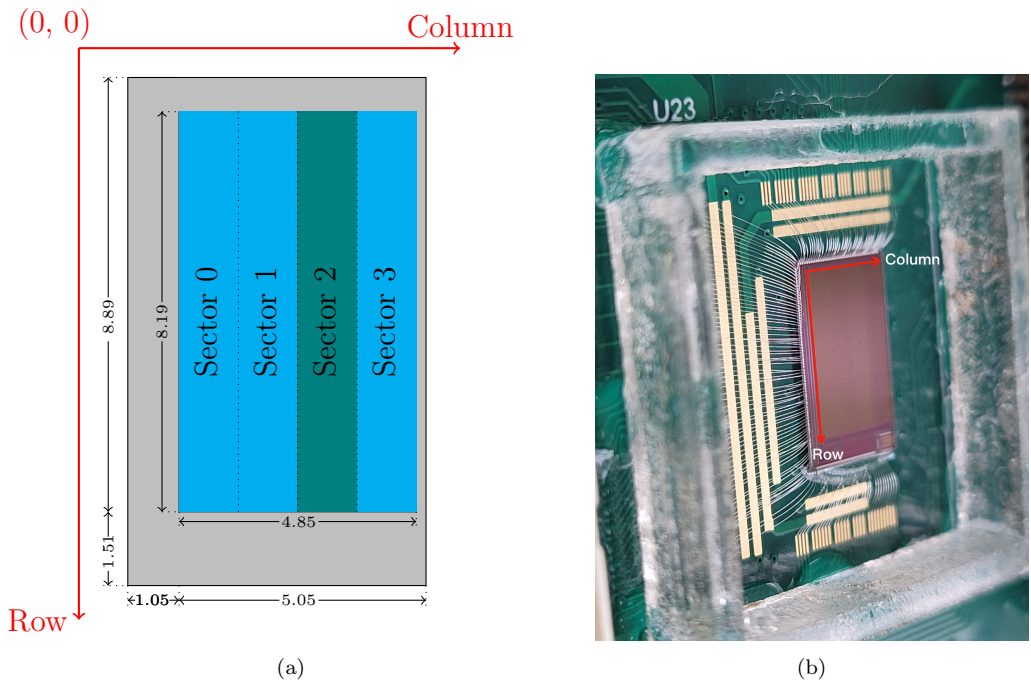

The JadePix-3 pixel sensor is a prototype design for the CEPC vertex detector, emphasizing high spatial resolution and low power consumption~\cite{zhu2019vertex}. It incorporates various analog front-end and digital circuit designs. The thickness of the sensor is \SI{300}{\um}. Its pixel matrix is divided into four distinct sectors.
Each sector comprises 512 rows and 48 columns of pixels. 
Using rolling shutter action, one row is activated at a time, requiring \SI{192}{\ns} for transmitting hit information and resetting pixels. Scanning the entire matrix of 512 rows takes \SI{98.3}{\us}.
The pixel pitch measures $26.0\times16.0$~\si{\um^2} for sectors 0, 1, and 3 and $23.11\times16.0$~\si{\um^2} for sector 2. Figure~\ref{fig:jadepix3-layout} depicts the sensor's layout, featuring a sensor size of $6.1\times10.4$ \si{\mm^2} and a pixel matrix area of $4.85\times8.19$~\si{\mm^2}.

The sensor's functions and performance have undergone extensive laboratory testing. Key parameters, such as the minimal threshold, power consumption, noise hit rate, and position resolution, have been thoroughly characterized\cite{dong2023design}.
An infra-red laser beam test system was used to assess the single-point resolution, giving hint on the expected spatial resolution in the range of $pitch/2\sqrt{12}$ and $pitch/\sqrt{12}$.
With the sensor's small pitch, particularly in the row direction, there is potential to construct a high-resolution beam telescope using this sensor. Our objective is to develop a high-precision measurement tool for testing the JadePix-3 itself and the subsequent series of chips through beam tests.

\FloatBarrier
\section{The JadePix-3 Telescope Implementation}
\subsection{Telescope setup}
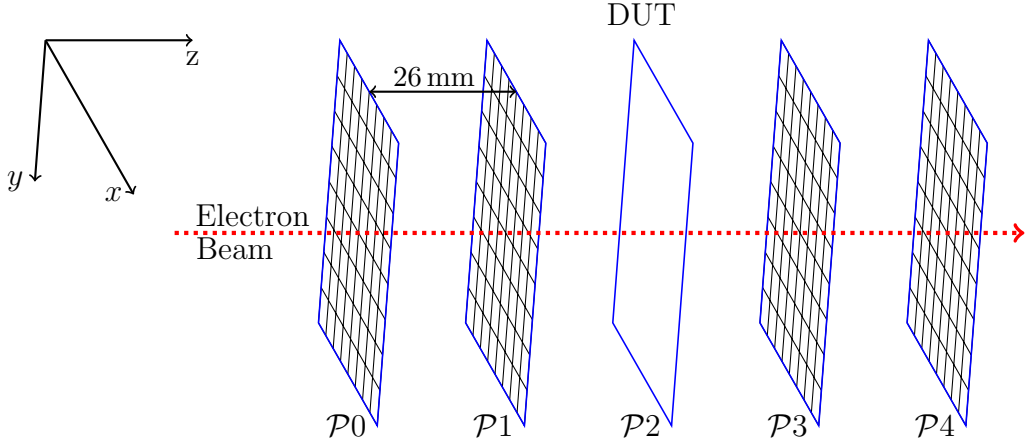
\begin{figure}[htbp]
    \centering
    \begin{tikzpicture}[3d view={20}{12},rotate=90]
        \foreach \Z in {0,...,4}
            {
                \begin{scope}[canvas is xy plane at z=2*-\Z,transform shape]
                    \ifnum \Z=2
                    \else
                        \draw (0,0) grid[step={pow(2,-1)}] (4,4);
                    \fi
                    \draw[semithick, blue] (0,0) coordinate(p\Z) rectangle (4,4);
                \end{scope}
                \path (p\Z) node[left]{$\mathcal{P}{\Z}$};
            }
        \draw[to-to, thick] (4,2,0) -- (4,2,-2);
        \node[] at (4,2.6,-1) {\small \SI{26}{\mm}};
        \draw[-to,red,dotted,ultra thick] (2,2,2.5) -- (2,2,-9.05);
        
        \node[] at (2, 2.7, 1.3) {Electron};
        \node[] at (2, 1.4, 1.8) {Beam};
        
        \node[] at (4, 5, -4.3) {DUT};
        
        \draw[thick,->]  (4,4,4) -- (4,-2,4) node[anchor= east] {\(x\)};
        \draw[thick,->] (4,4,4) -- (2,4,4) node[anchor= east] {\(y\)};
        \draw[thick,->]  (4,4,4) -- (4,4,2) node[anchor= north] {z};

    \end{tikzpicture}
    \caption{JadePix-3 telescope, composed of five parallel planes, each spaced \SI{26}{\mm} apart. P0, P1, P3, and P4 serve as reference planes, whereas P2 is defined as the DUT plane.}
    \label{fig:telescope-layout}
    
\end{figure}

\begin{figure}[htbp]
    \centering
    \begin{subfigure}[t]{0.31\textwidth}
        \centering
        \includegraphics[width=\textwidth]{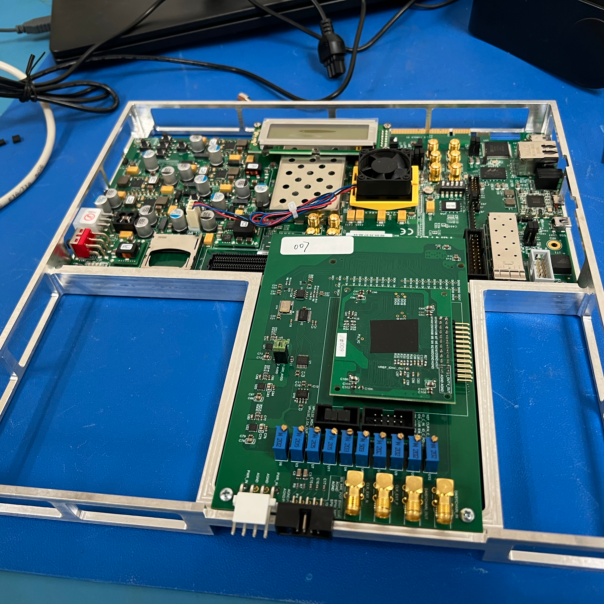}
        \caption{}
        \label{fig:single-plane}
    \end{subfigure}
    \hfill
    \begin{subfigure}[t]{0.31\textwidth}
        \centering
        \includegraphics[width=\textwidth]{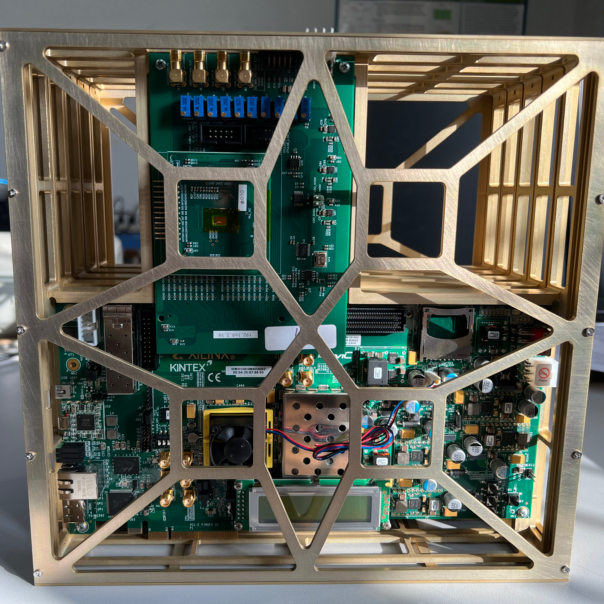}
        \caption{}
        \label{fig:front-view}
    \end{subfigure}
    \hfill
    \begin{subfigure}[t]{0.31\textwidth}
        \centering
        \includegraphics[width=\textwidth]{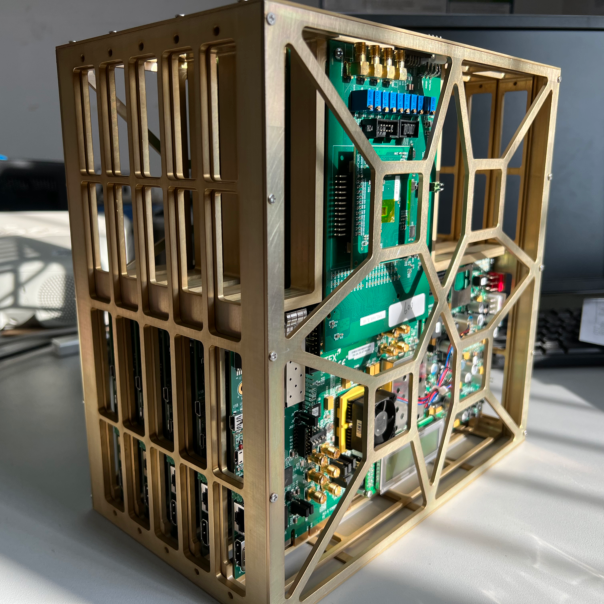}
        \caption{}
        \label{fig:side-view}
    \end{subfigure}
    \caption{Photographs of the JadePix-3 telescope. (a) Single telescope plane. (b) Front view. (c) Side view.}
    \label{fig:metal-frame}
    
\end{figure}

As depicted in Figure~\ref{fig:telescope-layout}, the JadePix-3 telescope comprises five planes, labeled P0, P1, P2, P3, and P4, with P2 serving as the device under test (DUT). The distance between the planes is fixed to \SI{26}{\mm}. In Figure~\ref{fig:metal-frame}, a compact and portable frame made of magnesium aluminum metal is employed to support and protect the telescope. Each telescope plane includes a commercial FPGA evaluation board (KC705), a sensor bonding board, and an FMC adaptor facilitating communication between the sensor and the FPGA. The FPGA board is used for configuring the parameters of the sensor, controlling its readout process, and transmitting the sensor data to the PC via the Ethernet.

The global coordinate system is established as a right-handed Cartesian system, with the positive z-axis aligned with the beam direction, and the origin defined within the plane P2.

\subsection{Design for synchronous readout of multiple chips}
A data acquisition (DAQ) system, using the IPbus framework, was developed~\cite{dong2021daq}. The IPbus framework, which employs the User Datagram Protocol for network communication, offers excellent scalability~\cite{larrea2015ipbus}. This scalability was a key factor in selecting IPbus, as it facilitated the seamless expansion to a multi-chip readout system.

\begin{figure}[htbp]
    \centering
    \includegraphics[width=\textwidth]{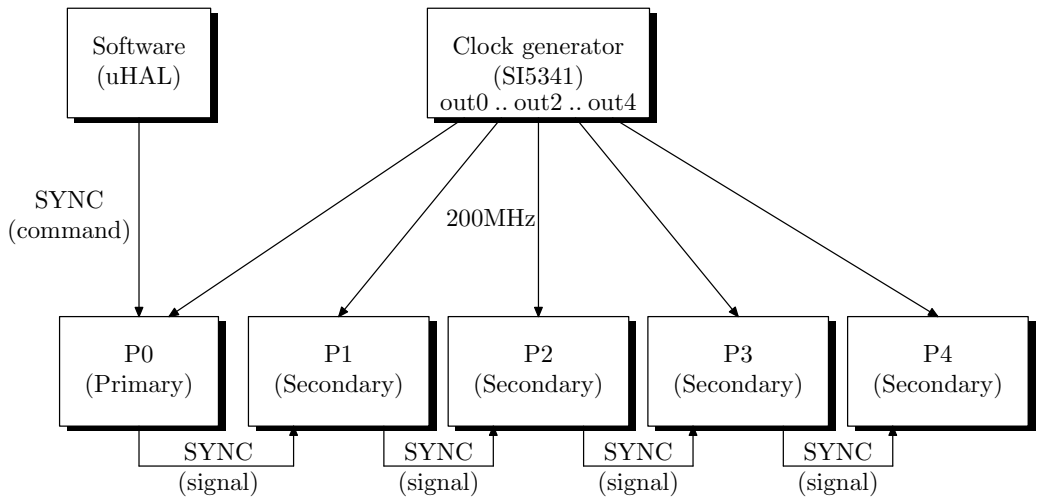}
    \caption{Clock distribution and synchronized start signal link in the readout design. A software command using the IPbus protocol is sent to the primary plane P0, and a synchronous start command (SYNC) signal is transmitted via a daisy-chain link to the next secondary plane sequentially.}
    \label{fig:daq}
\end{figure}

We used the SI5341-D-EVB, a commercial clock generator evaluation board, to provide a synchronous system clock for each telescope plane, as depicted in Figure~\ref{fig:daq}. This configuration ensures that the telescope planes operate within the same clock domain. To synchronize the start of time in each telescope plane, a daisy-chain signal link was devised. The synchronous start command (SYNC) is initially issued via an end-user Application Program Interface (API) based on the micro Hardware Access Library (uHAL). Upon receiving the SYNC command, the primary plane P0 generates a SYNC signal and sequentially transmits it to the next secondary plane. An external trigger is not necessary, as the sensors are read out continuously for all the pixel hits. The delay in the daisy-chain link is less than \SI{100}{\ns}, a duration negligible compared with the sensor's integration time of \SI{98.3}{\us} for each readout frame. This approach guarantees that all planes start operations in a synchronized way.

\FloatBarrier
\section{Experimental Setup}
\begin{figure}[htbp]
    \centering
    \includegraphics[width=.9\textwidth]{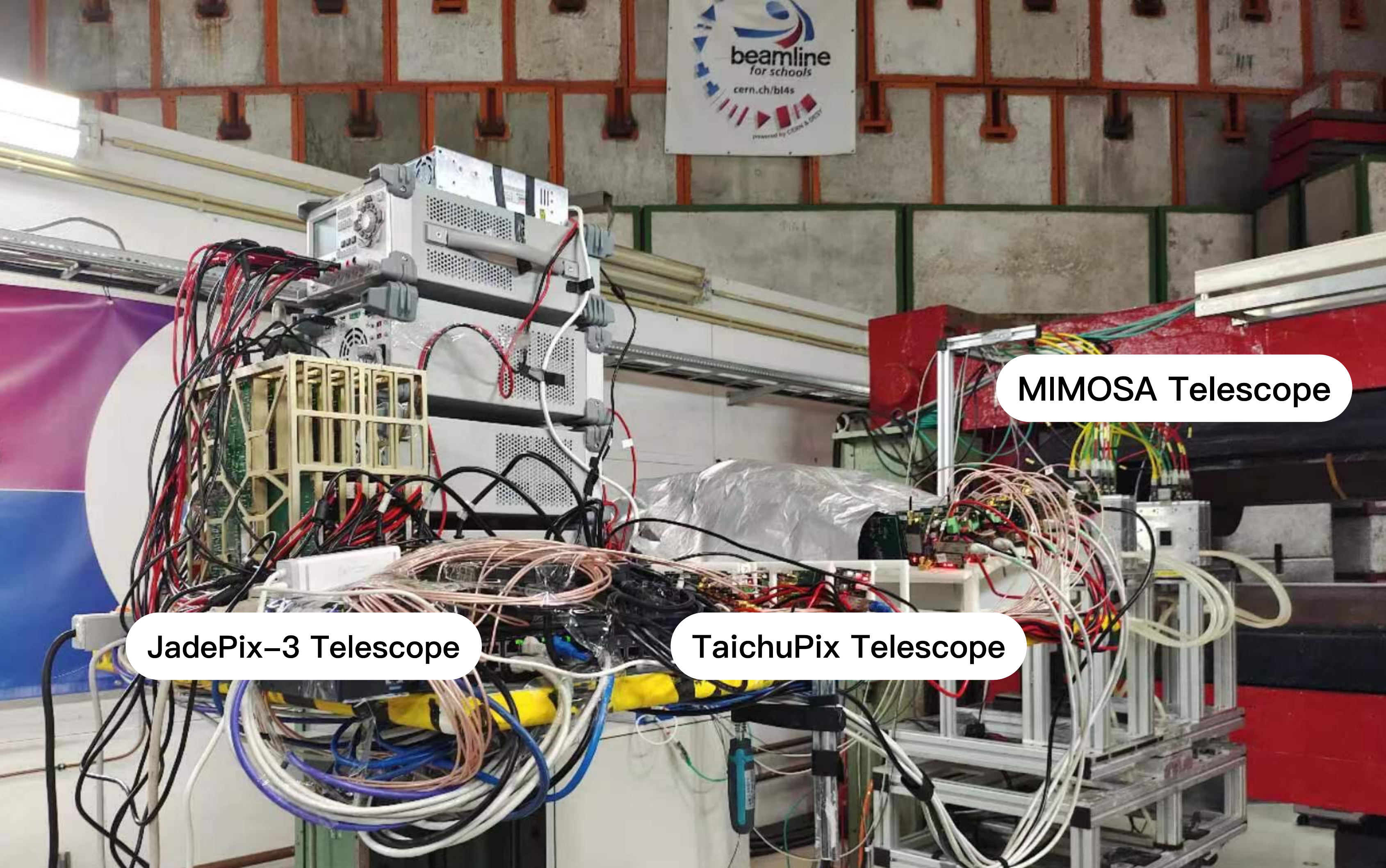}
    \caption{Experimental setup at DESY TB21. The JadePix-3 telescope was positioned downstream of the beamline, following the MIMOSA and TaichuPix telescopes. The JadePix-3 telescope shared the beam time and collected data independently.}
    \label{fig:experimental-set-up}
\end{figure}

The DESY-II test beam facility provides electron/positron beams with user-selectable momenta ranging from 1 to \SI{6}{GeV/c}, and a divergence of approximately \SI{1}{mrad}~\cite{diener2019desy}. Beam test data were collected at TB21, one of the three independent beamlines at the facility. The JadePix-3 telescope was positioned downstream of the beamline, behind the MIMOSA~\cite{jansen2016performance, hu2010first} and the TaichuPix~\cite{wu2024beam} telescopes. Figure~\ref{fig:experimental-set-up} illustrates the experimental setup. The JadePix-3 telescope shared the beam time and collected data independently.

\FloatBarrier
\section{Data Analysis and Results Discussion}
The Corryvreckan framework~\cite{dannheim2021corryvreckan} is used for offline data analysis. The standard procedure for offline analysis involves decoding raw data, clustering, alignment, and track fitting. Subsequently, the performance of the Telescope and DUT is analyzed based on this foundation. 

A dedicated event loader module, EventLoaderJadepix3, was developed to import binary hit information from the raw data files of each JadePix-3 sensor. The Metronome module was employed to segment the data stream into regular time frames of a specified length. The time frame was set to \SI{98.316}{\us}, corresponding to the rolling shutter frame readout time. Standard Corryvreckan modules, including clustering, tracking, and alignment, were used for precise alignment. Two track reconstruction methods were employed: straight-line and general broken lines (GBL). The GBL served as the primary reconstruction algorithm, considering the effects of multiple scattering and providing more accurate analytical results. The straight-line method was used for analyzing track angles.

To simplify the analysis, data from the left half (sectors 0 and 1) of the pixel matrix were chosen, whereas data from the right half (sectors 2 and 3) were excluded because of the different pitch in sector 2. Consequently, the pixel pitch related to the analytical results corresponds to a size of $26\times16$~\si{\um^2}.

\subsection{Cluster size distribution}
The Clustering4D module in Corryvreckan is used to cluster pixel hits that have coinciding timestamps. The cluster is identified within a dynamic radius, up to the maximum extent at which a split cluster can be recognized. Given that JadePix-3 sensors only provide 1-bit binary information, the position of a cluster is determined by the geometric center of adjacent pixel hits.

\begin{figure}[htbp]
    \centering
    \centering
    \begin{subfigure}[b]{0.495\textwidth}
        \centering
        \includegraphics[width=\textwidth]{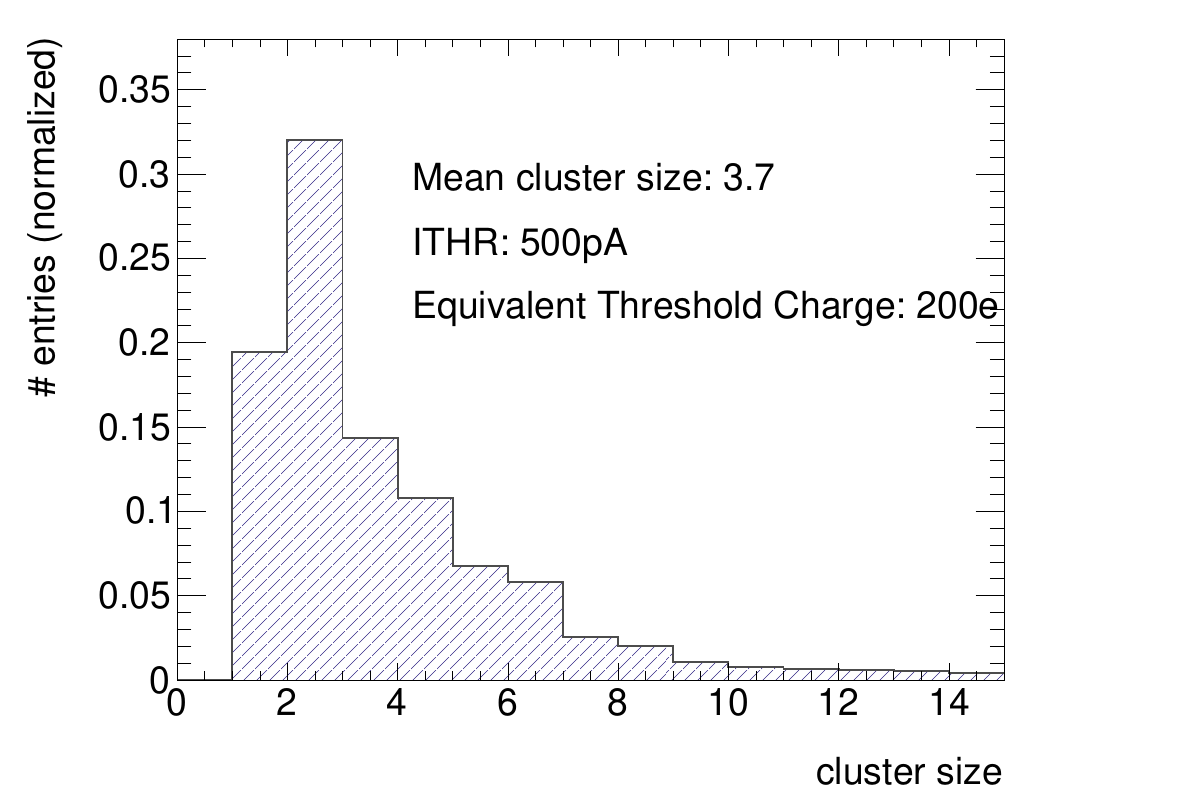} 
        \caption{}
        \label{fig:cluster_0}
    \end{subfigure}
    \hfill
    \begin{subfigure}[b]{0.495\textwidth}
        \centering
        \includegraphics[width=\textwidth]{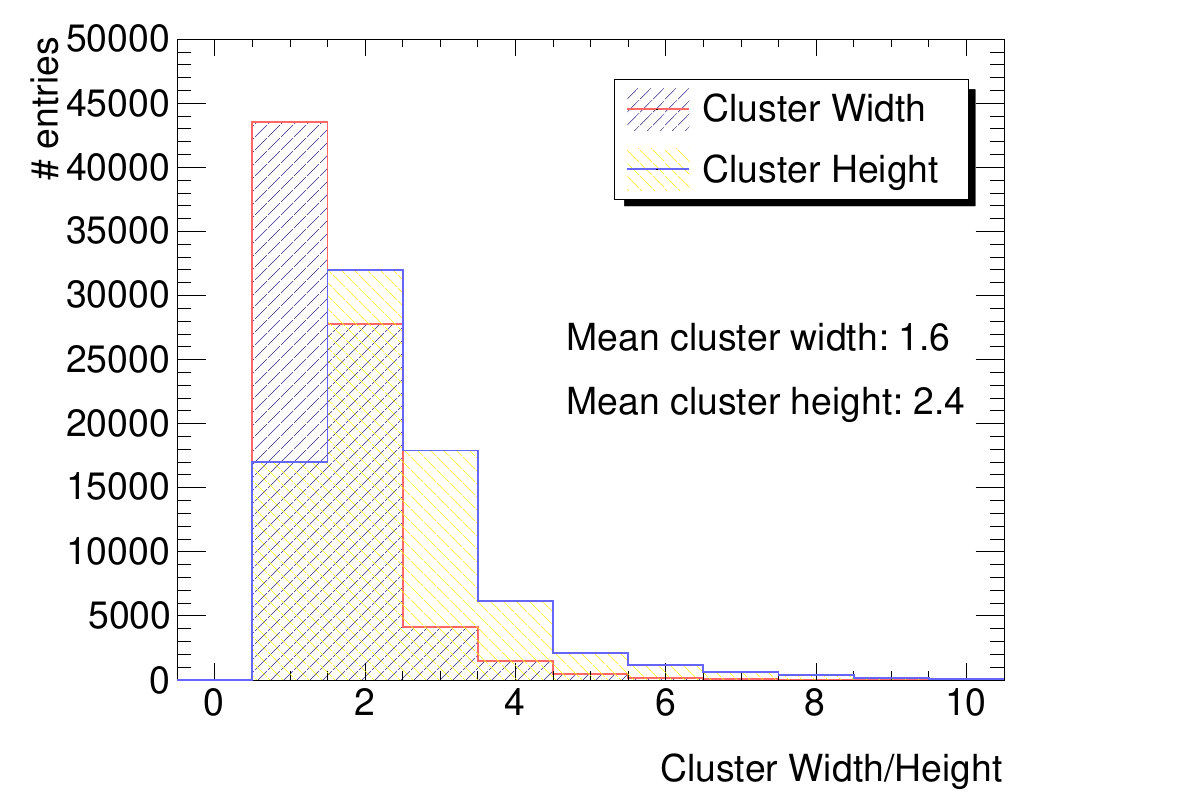} 
        \caption{}
        \label{fig:cluster_1}
    \end{subfigure}
    \caption{(a) Cluster size distribution of the DUT. (b) Distribution of cluster width and cluster height of the DUT. Experiment parameters: beam energy = \SI{5.8}{\GeV}, threshold = \SI{200}{\electron}.}
    \label{fig:cluster}
\end{figure}

Figure~\ref{fig:cluster_0} displays the cluster size (CS) distribution of the DUT with the threshold set to \SI{200}{\electron}. The average CS is found to be 3.7, with average cluster width and height (projected in the \(x\) and \(y\) directions) measured at 1.6 and 2.4, respectively, as illustrated in Figure~\ref{fig:cluster_1}. The difference in cluster in width and height result from the asymmetric dimensions of the pixels. The normalized CS distribution, as shown in Figure~\ref{fig:cluster_0}, shows a gradual decrease for CSs larger than 2. CSs smaller than 4 account for \SI{76}{\percent} of the total number, with those smaller than 9 constituting \SI{95}{\percent}.

\begin{figure}[htbp]
    \centering
    \begin{subfigure}[b]{0.49\textwidth}
        \centering
        \includegraphics[width=\textwidth]{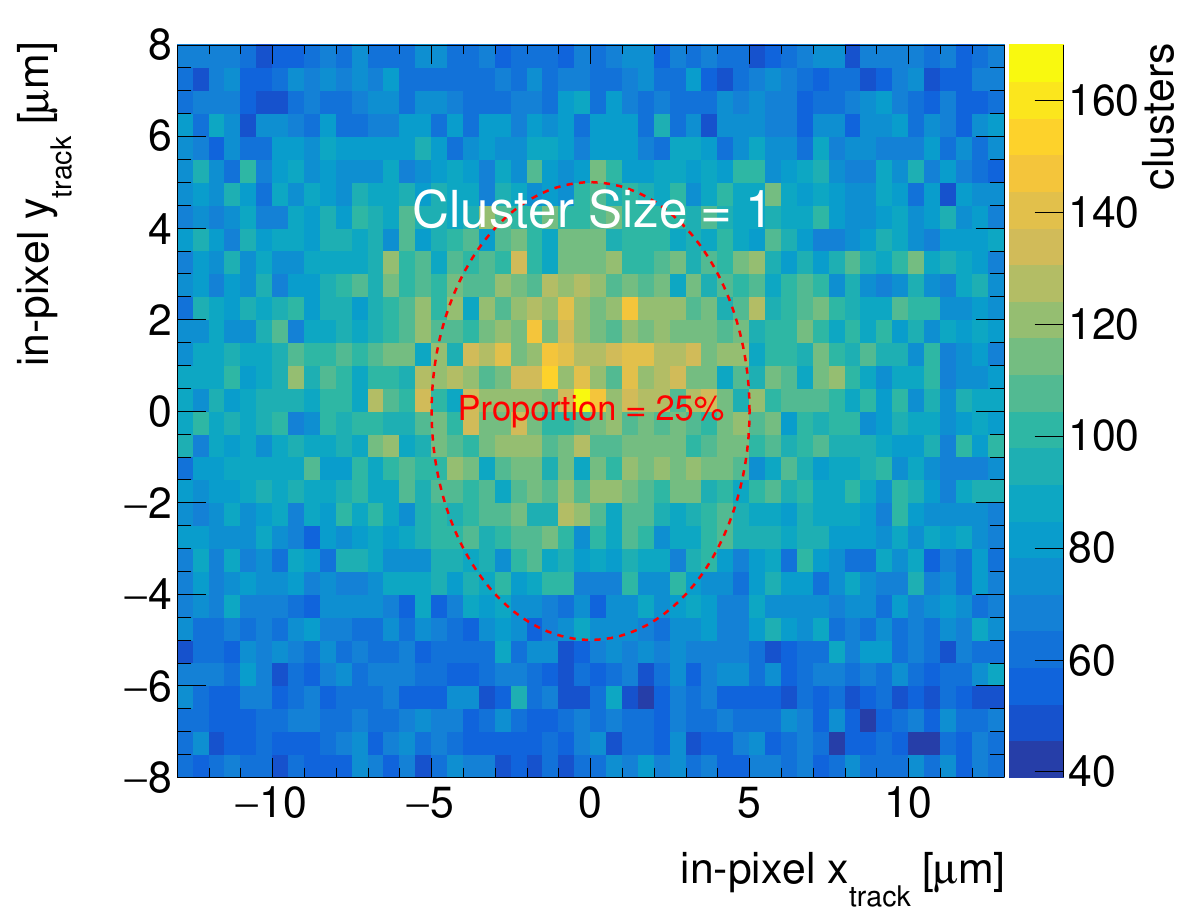} 
        \caption{}
        \label{fig:cs1_map}
    \end{subfigure}
    \hfill
    \begin{subfigure}[b]{0.49\textwidth}
        \centering
        \includegraphics[width=\textwidth]{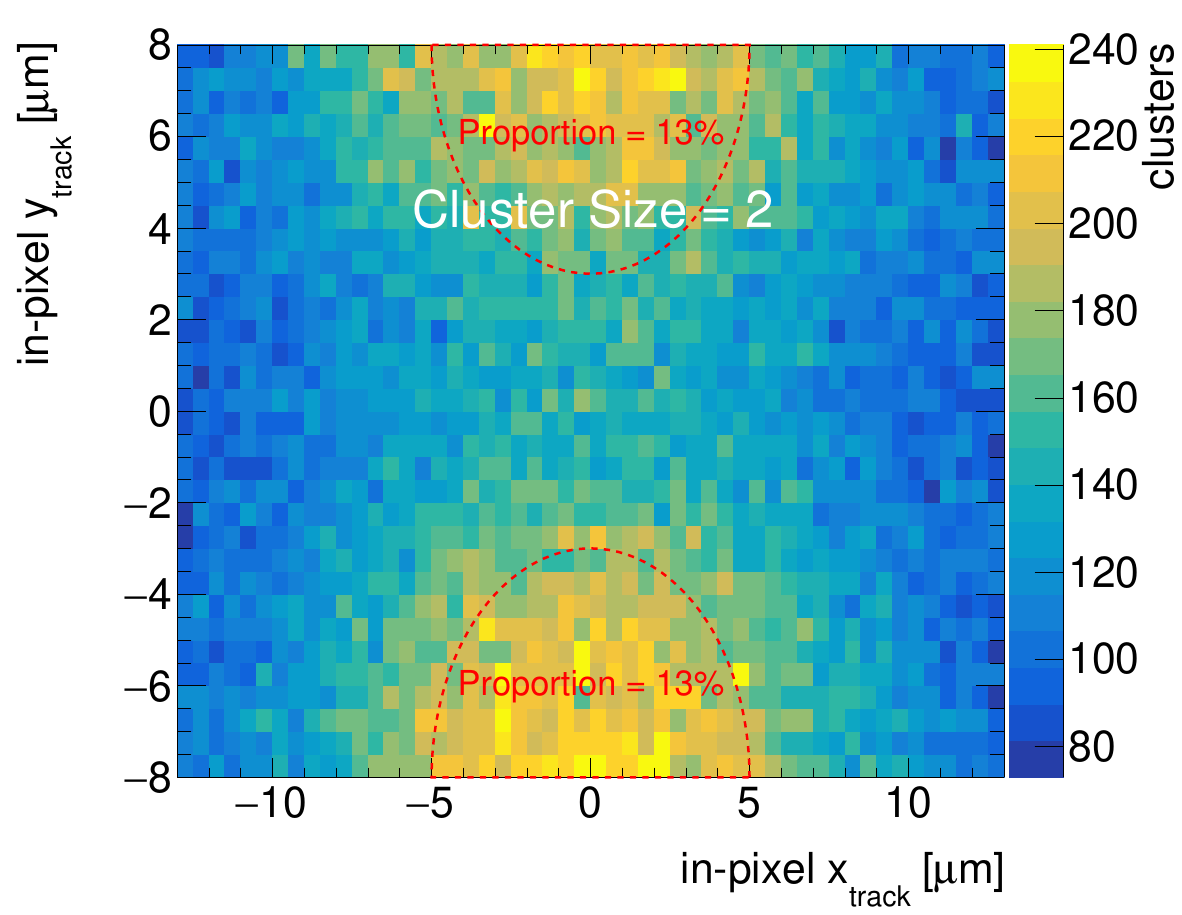} 
        \caption{}
        \label{fig:cs2_map}
    \end{subfigure}
    \centering
    \begin{subfigure}[b]{0.49\textwidth}
        \centering
        \includegraphics[width=\textwidth]{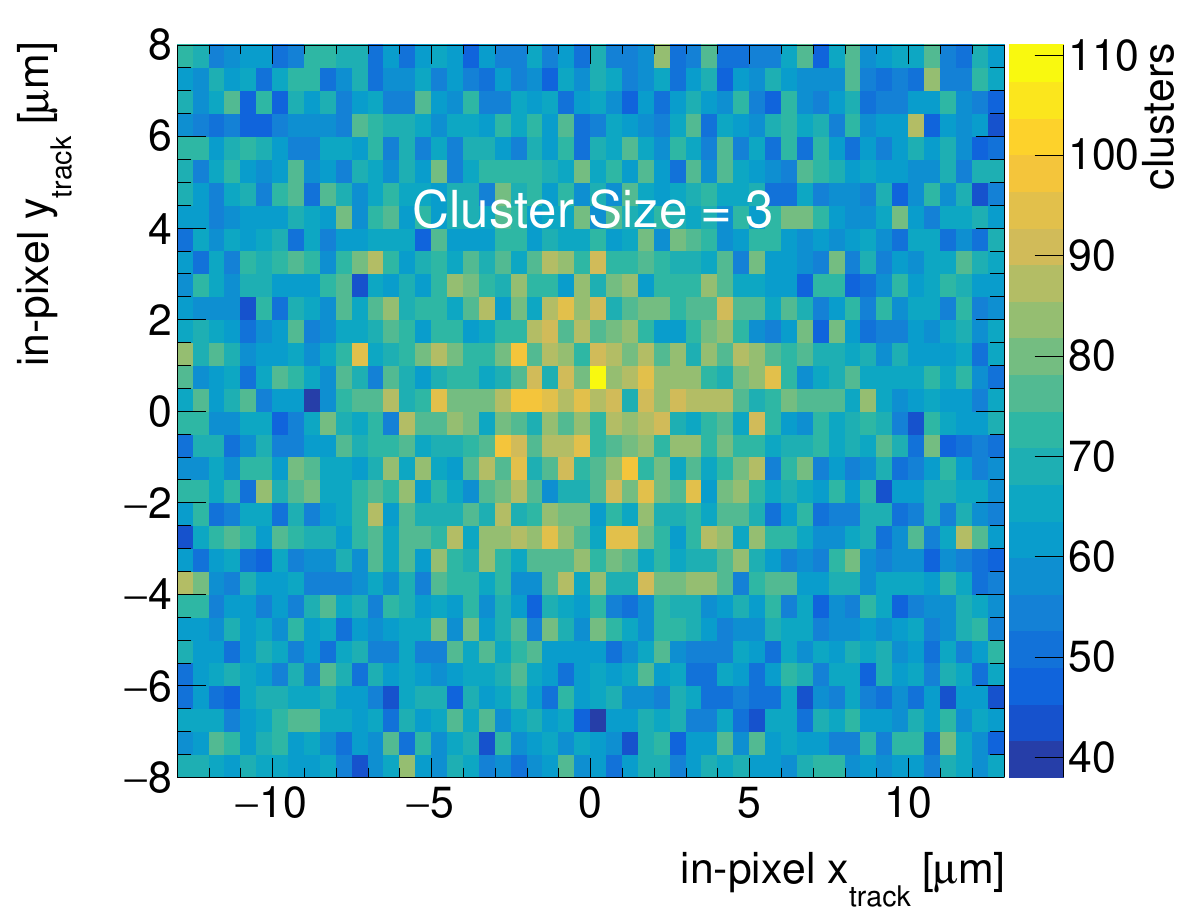} 
        \caption{}
        \label{fig:cs3_map}
    \end{subfigure}
    \hfill
    \begin{subfigure}[b]{0.49\textwidth}
        \centering
        \includegraphics[width=\textwidth]{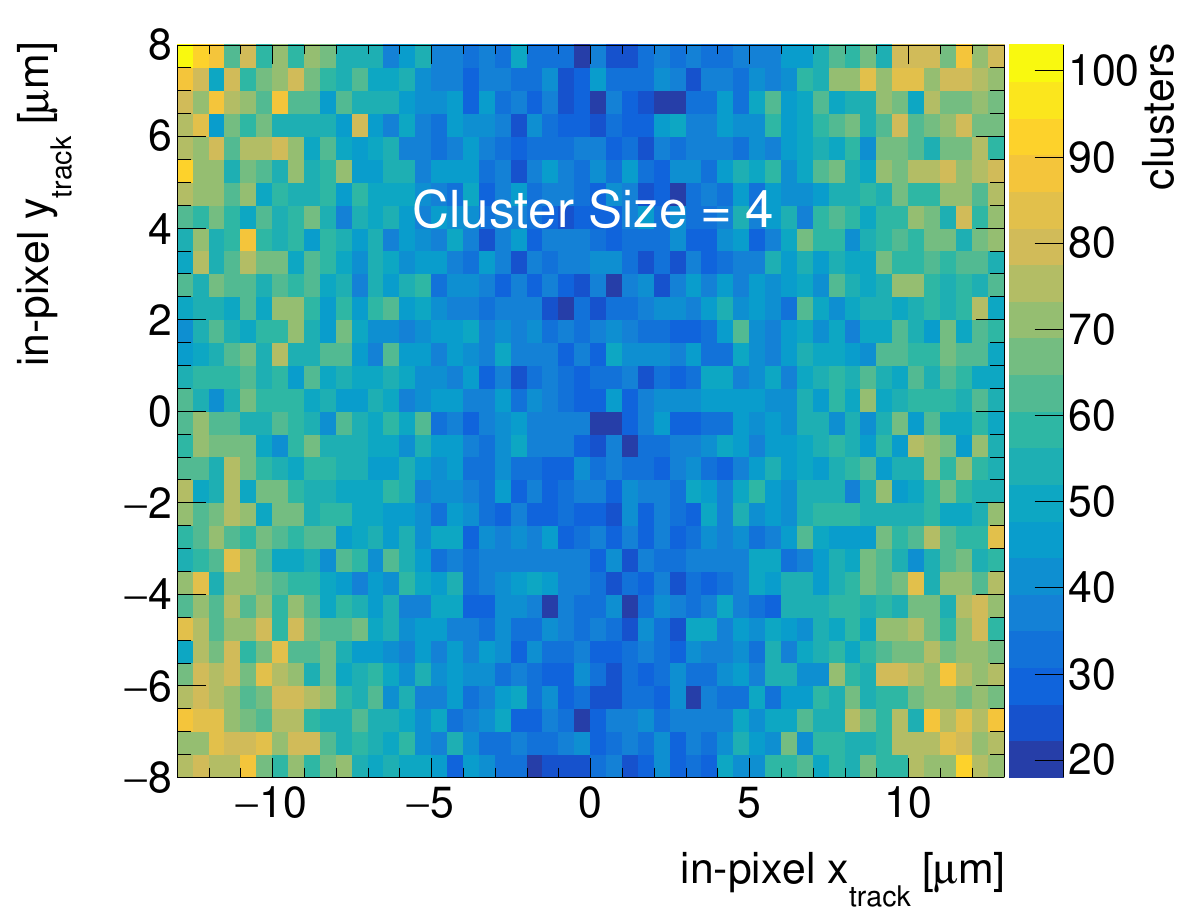} 
        \caption{}
        \label{fig:cs4_map}
    \end{subfigure}
    \caption{Unbiased intra-pixel track distribution at the DUT for cluster sizes ranging from 1 to 4. Experiment parameters: beam energy = \SI{5.8}{\GeV}, threshold = \SI{200}{\electron}.}
    \label{fig:cs_map}
\end{figure}

Figure~\ref{fig:cs_map} illustrates the intra-pixel track distribution with CSs of 1, 2, 3, and 4 pixels. When the CS is 1, the particle's incident position is more likely to be at the center of the pixel, as shown in Figure~\ref{fig:cs1_map}. The proportion of events at the center of the diode within a radius of less than approximately \SI{5}{\um} is \SI{25}{\percent}. In contrast, when the hit position is at the four sides of a pixel, multiple pixels are more likely to be fired owing to the charge-sharing effect. This effect is especially pronounced in the middle of the long sides with a CS of 2, as depicted in Figure~\ref{fig:cs2_map}, and at the corners with a CS of 4, as shown in Figure~\ref{fig:cs4_map}. For cluster size 2, the proportion of events in the semicircular area with a radius of \SI{5}{\um} along the longer side is \SI{13}{\percent}. For comparison, the results obtained on a different pixel pitch of $18.4\times18.4$~\si{\um^2} can be found in reference~\cite{jansen2016resolution}.

\begin{figure}[htbp]
    \centering
    \begin{subfigure}[b]{0.495\textwidth}
        \centering
        \includegraphics[width=\textwidth]{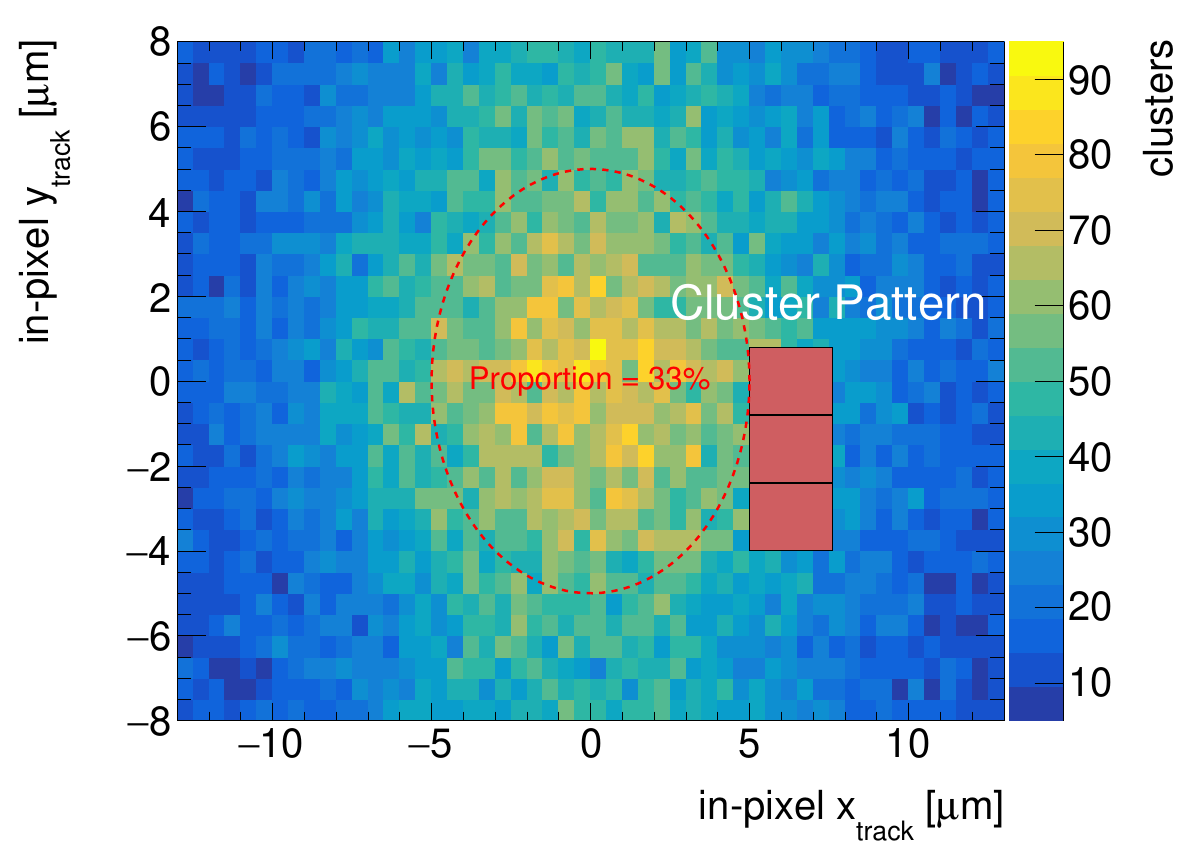} 
        \caption{}
        \label{fig:cs3_p1}
    \end{subfigure}
    \hfill
    \begin{subfigure}[b]{0.495\textwidth}
        \centering
        \includegraphics[width=\textwidth]{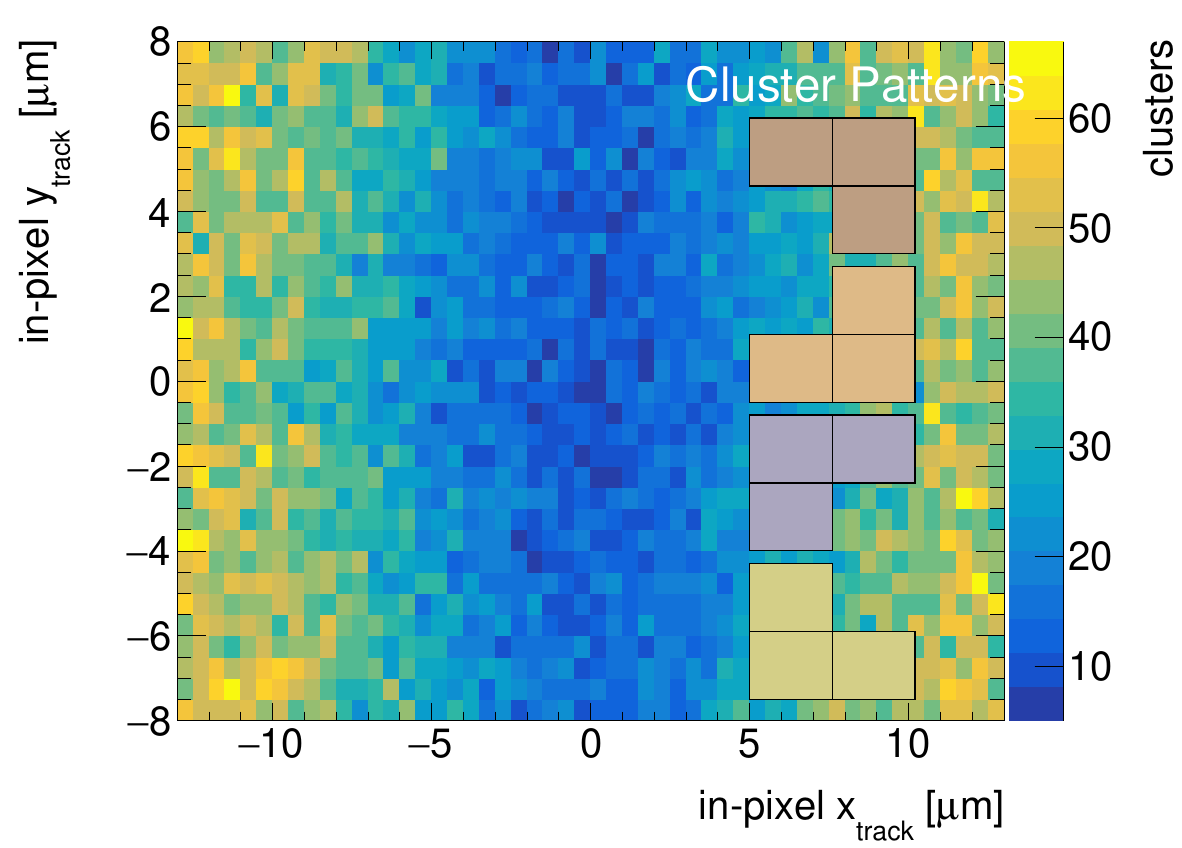} 
        \caption{}
        \label{fig:cs3_p2}
    \end{subfigure}
    \caption{Intra-pixel track distribution for two categories of cluster patterns with CS = 3. (a) Cluster Pattern 1 with three pixels lined consecutively in the \(y\) direction (b) Cluster Pattern 2 with three pixels spanned in both the \(x\) and \(y\) directions. Experiment parameters: beam energy = \SI{5.8}{\GeV}, threshold = \SI{200}{\electron}.}
    \label{fig:cs3_pattern}
\end{figure}

To further explore the factors influencing the track distribution for a CS of 3, as demonstrated in Figure~\ref{fig:cs3_map}, we categorized various cluster patterns and plotted their intra-pixel track distribution, as depicted in Figure~\ref{fig:cs3_pattern}. Two distinct categories of patterns were identified: Pattern 1, as shown in Figure~\ref{fig:cs3_p1}, is characterized by three pixels lined consecutively in the \(y\) direction; and Pattern 2, demonstrated in Figure~\ref{fig:cs3_p2}, features three pixels spanned in both the \(x\) and \(y\) directions. 

The occurrence rates for Patterns 1 and 2 are \SI{53.3}{\percent} and \SI{46.3}{\percent}, respectively. Considering the longer \(x\) dimension of the pixels, horizontal clusters of three consecutive pixels are rare, occurring at a rate of only \SI{0.4}{\percent}. In Pattern 1, the firing of the central pixel tends to have the pixels immediately above and below it fired as well. The proportion of events at the center of the diode within a radius of less than approximately \SI{5}{\um} is \SI{33}{\percent}. 
In Pattern 2, firing occurs more readily at the short side of pixels, leading to adjacent pixels firing in both the \(x\) and \(y\) directions.

\FloatBarrier
\subsection{Alignment and tracking}
According to the guidelines outlined in the Corryvreckan manual~\cite{kroger2019user}, the alignment procedure comprises the following steps: (1) pre-align the telescope planes, (2) finely align the telescope planes, (3) pre-align the DUT plane, and (4) finely align the DUT plane. For steps 1 and 2, the alignment requires each telescope plane to have a concurrent cluster.

\begin{figure}[ht]
    \centering
    \centering
    \begin{subfigure}[b]{0.495\textwidth}
        \centering
        \includegraphics[width=\textwidth]{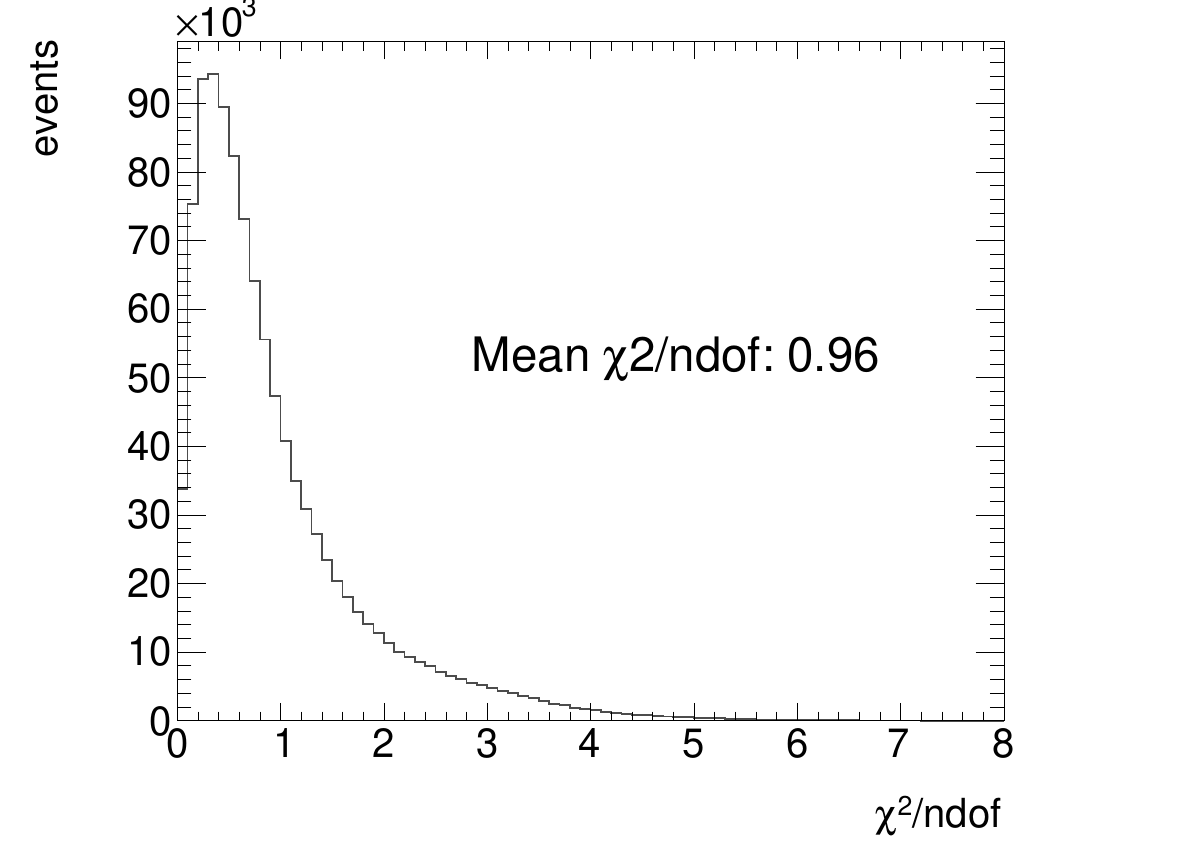} 
        \caption{}
        \label{fig:track_0}
    \end{subfigure}
    \hfill
    \begin{subfigure}[b]{0.495\textwidth}
        \centering
        \includegraphics[width=\textwidth]{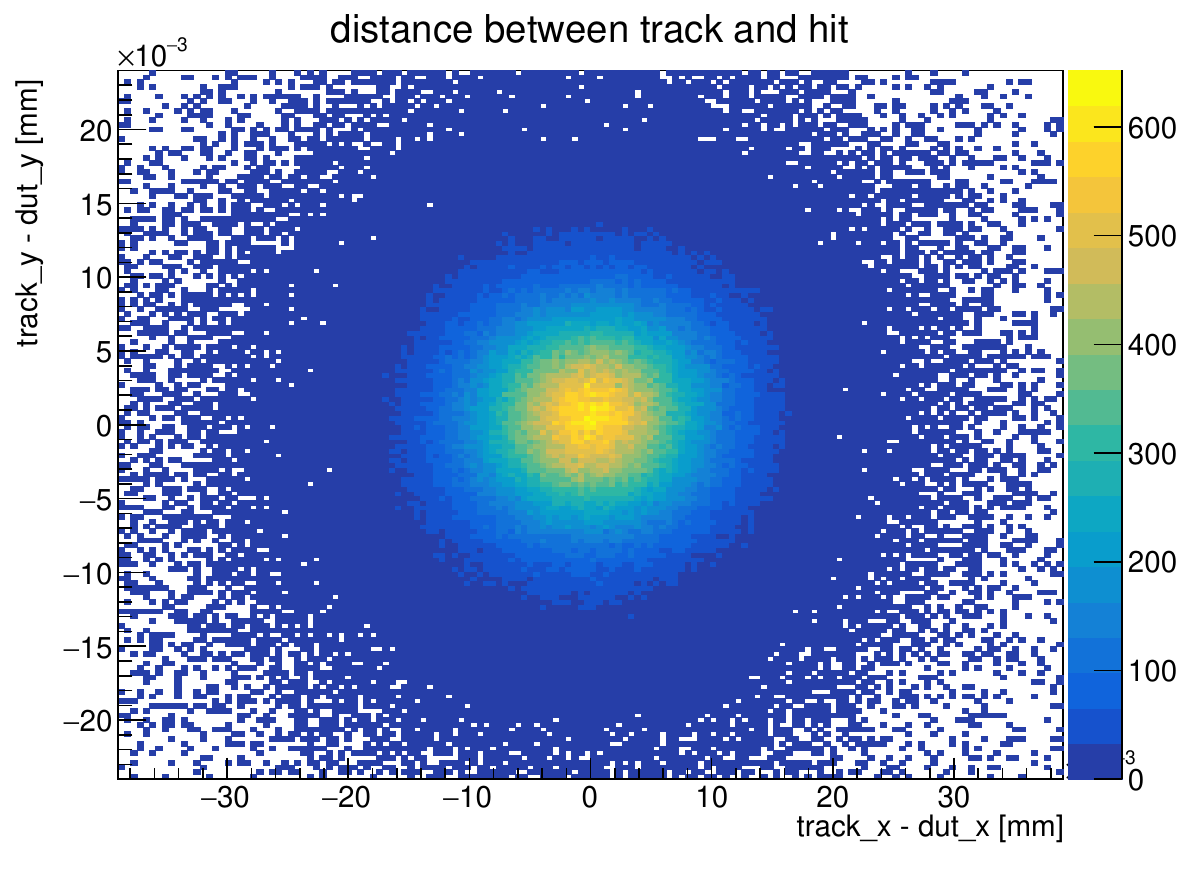} 
        \caption{}
        \label{fig:track_1}
    \end{subfigure}
    \caption{Assessment of alignment performance through (a) the distribution of $\chi^2/{n_{dof}}$ of track fitting and (b) the distribution of the distance between the track and the hit. Experiment parameters: beam energy = \SI{5.8}{\GeV}, threshold = \SI{200}{\electron}.}
    \label{fig:track}
\end{figure}

The alignment performance can be evaluated by examining the $\chi^2/{n_{dof}}$ (chi-square per degree of freedom) of track fitting and the distance distribution between the track and the hit. Figure~\ref{fig:track_0} displays the $\chi^2/{n_{dof}}$ distribution, with a mean value of 0.96. The notably low mean value indicates a close alignment between the fitted tracks and the cluster positions. In addition, Figure~\ref{fig:track_1} illustrates the distribution of distances between the track and hit, revealing minimal distances in both the \(x\) and \(y\) directions, with symmetry about the origin in both directions. These results collectively indicate precise and reliable tracking performance.

\FloatBarrier
\subsection{Residual and spatial resolution}
\begin{figure}[h]
    \centering
    \begin{subfigure}[b]{0.495\textwidth}
        \centering
        \includegraphics[width=\textwidth]{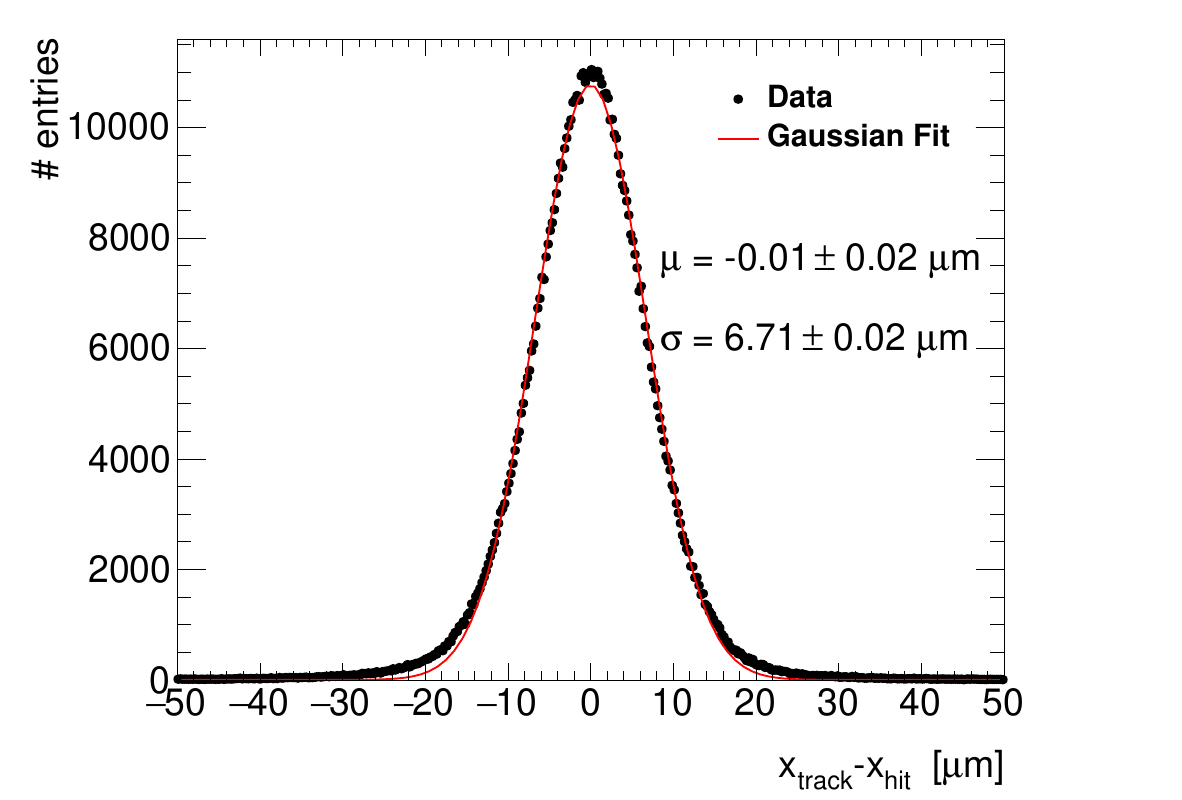} 
        \caption{}
        \label{fig:res_x}
    \end{subfigure}
    \hfill
    \begin{subfigure}[b]{0.495\textwidth}
        \centering
        \includegraphics[width=\textwidth]{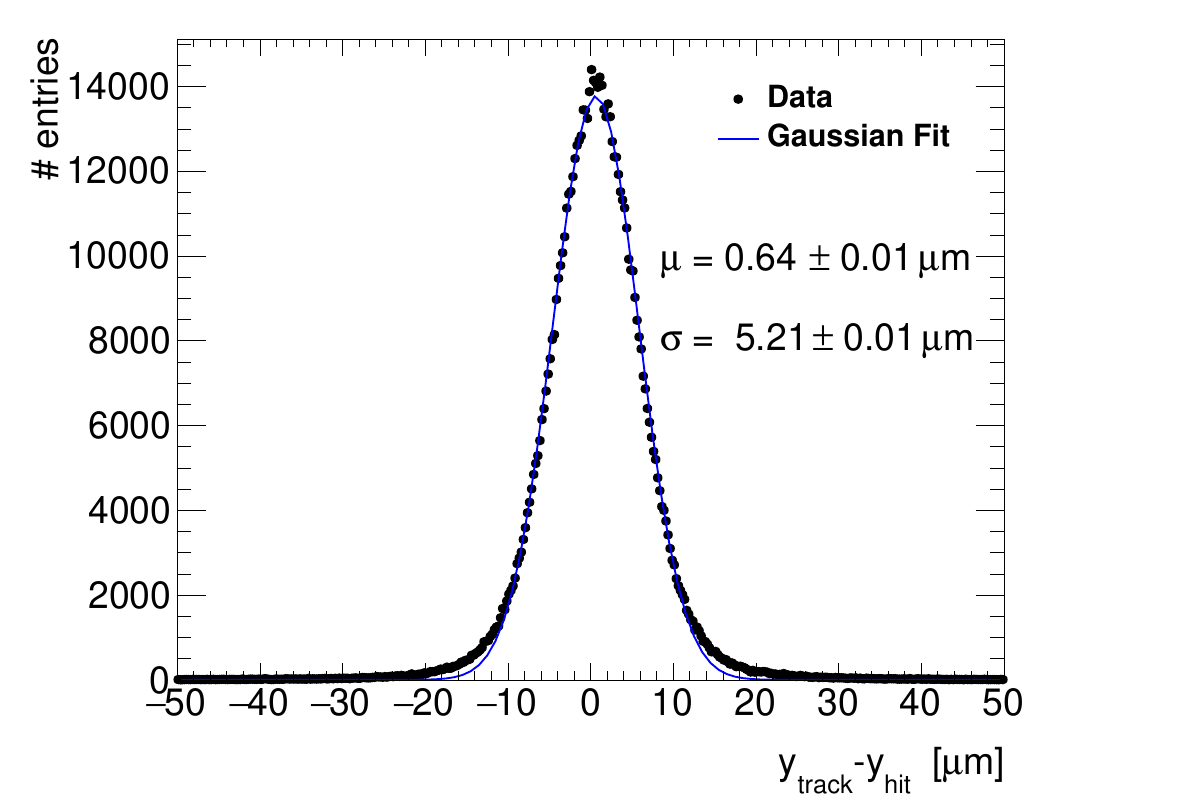} 
        \caption{}
        \label{fig:res_y}
    \end{subfigure}
    \caption{Unbiased residual distribution in the (a) \(x\) and (b) \(y\) directions, with the data represented by dots and the Gaussian fit shown in red/blue lines. Experiment parameters: beam energy = \SI{5.8}{\GeV}, threshold = \SI{200}{\electron}.}
    \label{fig:res_0}
\end{figure}

Using track fitting, hit prediction points corresponding to the DUT measurement points are obtained, enabling the determination of the hit position residual distribution on the DUT. The unbiased residual distributions in the  \(x\) and \(y\) directions are illustrated in Figure~\ref{fig:res_0}, where the red/blue line represents the Gaussian fit of the data. The $\sigma$ of the position residual distribution for the DUT in the \(x\) and \(y\) directions are 6.7 and \SI{5.2}{\um}, respectively.

To precisely determine the spatial resolution of the DUT plane, it is essential to remove the contribution of the telescope system, as described in Equation~\ref{eq:eq1}. As indicated in Equation~\ref{eq:eq2}, $\sigma_{tel}$ can be related to $\sigma_{plane}$ with a scaling factor \( k \)~\cite{bulgheroni2010results}, assuming that all planes have the same intrinsic resolution and that the reference planes are symmetrically positioned around the DUT at \( z_0 \). The factor \( k \) can be calculated using Equation~\ref{eq:eq3}, which considers the distance between each plane.

\begin{equation}
    \sigma_{meas}^2 = \sigma_{DUT}^2 + \sigma_{tel}^2
    \label{eq:eq1}
\end{equation}

\begin{equation}
    \sigma_{tel}^2 = k\sigma_{plane}^2
    \label{eq:eq2}
\end{equation}

\begin{equation}
    k=\frac{\sum_i^N z_i^2}{N\sum_i^N z_i^2-(\sum_i^Nz_i)^2}
    \label{eq:eq3}
\end{equation}

Using Equations~\ref{eq:eq1} and \ref{eq:eq2}, we can deduce Equations~\ref{eq:eq4} and \ref{eq:eq5}.

\begin{equation}
    \sigma_{DUT}^2 = \sigma_{plane}^2 = \frac{\sigma_{meas}^2 }{1+k}
    \label{eq:eq4}
\end{equation}

\begin{equation}
    \sigma_{tel}^2 = \frac{k}{1+k}\sigma_{meas}^2
    \label{eq:eq5}
\end{equation}

The scaling factor \( k \) for the JadePix-3 telescope has been calculated as 0.25. Consequently, the spatial resolution for the DUT is determined to be 6.0 and \SI{4.7}{\micro\meter} in the \(x\) and \(y\) directions, respectively. In addition, the resolution of the telescope is found to be 3.0 and \SI{2.3}{\micro\meter} in the \(x\) and \(y\) directions, respectively.

\begin{figure}[ht]
    \centering
    \begin{subfigure}[b]{0.495\textwidth}
        \centering
        \includegraphics[width=\textwidth]{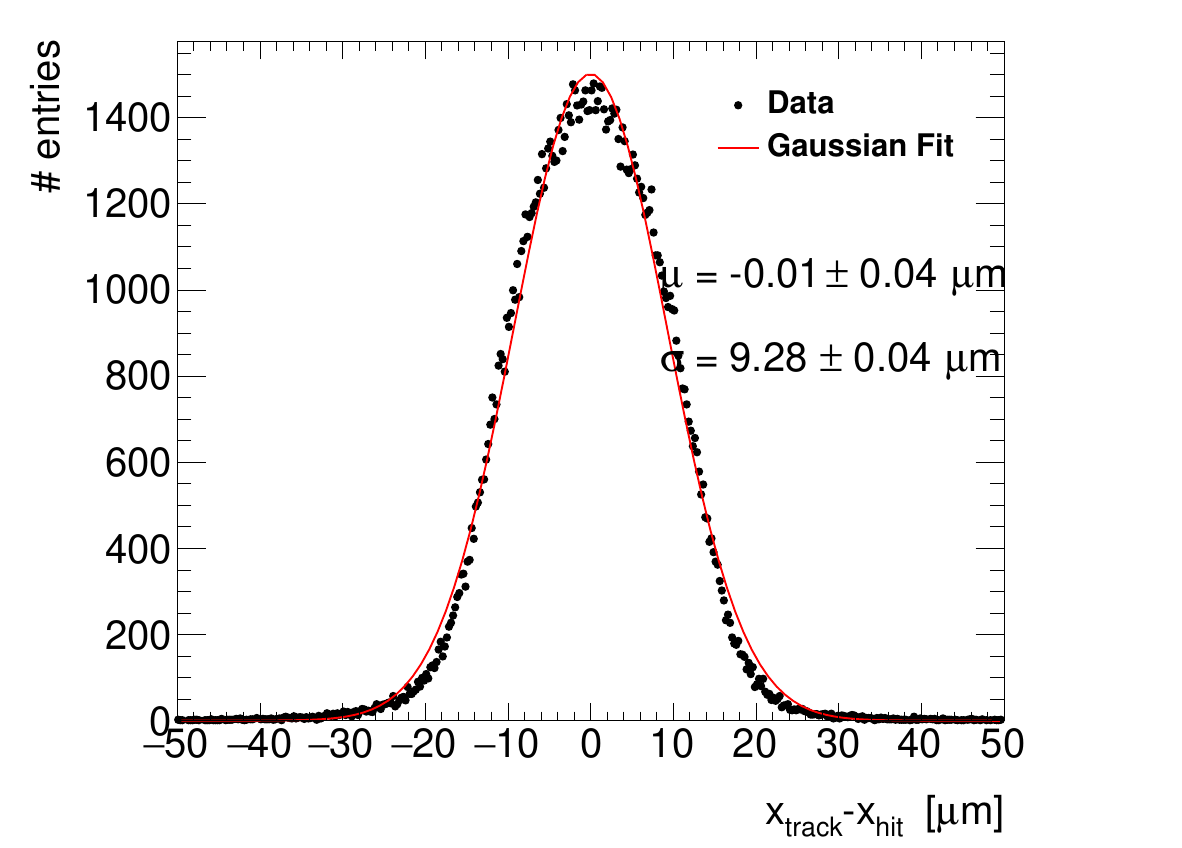} 
        \caption{}
        \label{fig:resX_cs1}
    \end{subfigure}
    \hfill
    \begin{subfigure}[b]{0.495\textwidth}
        \centering
        \includegraphics[width=\textwidth]{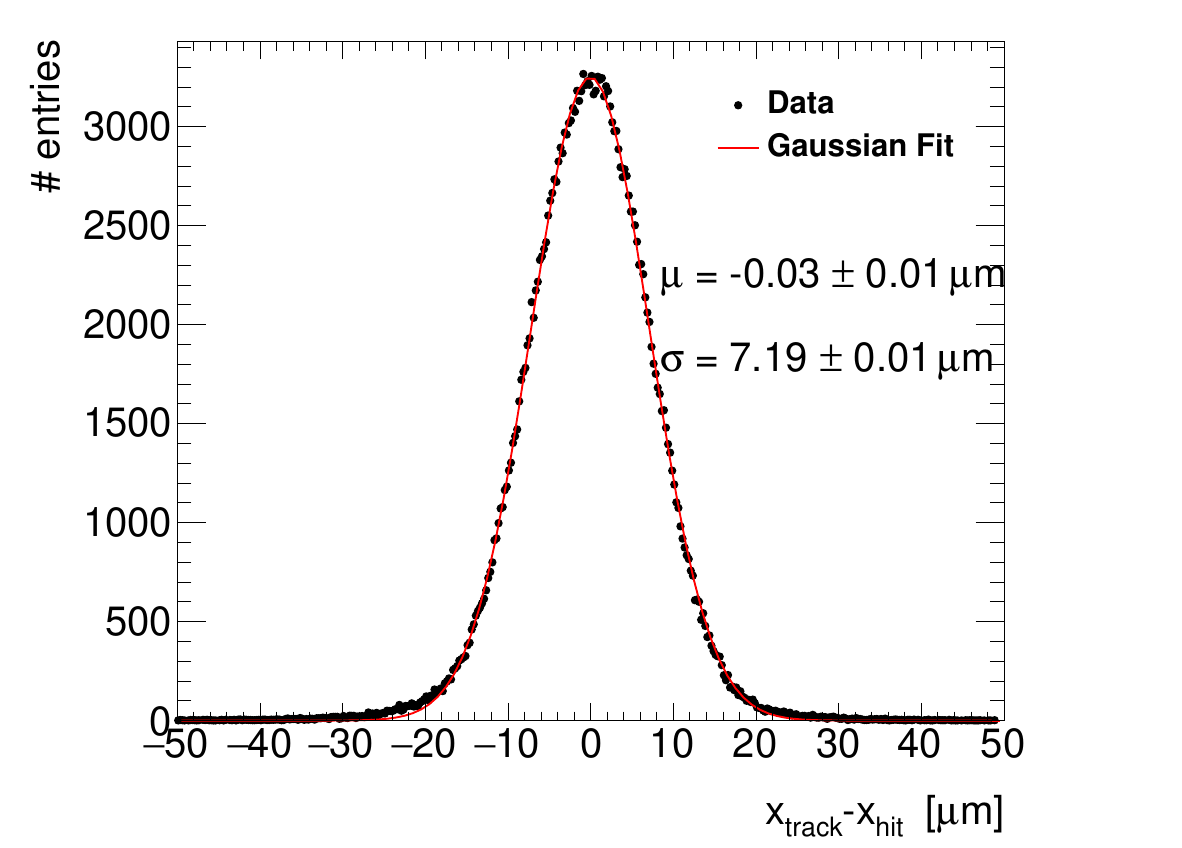} 
        \caption{}
        \label{fig:resX_cs2}
    \end{subfigure}
    \centering
    \begin{subfigure}[b]{0.495\textwidth}
        \centering
        \includegraphics[width=\textwidth]{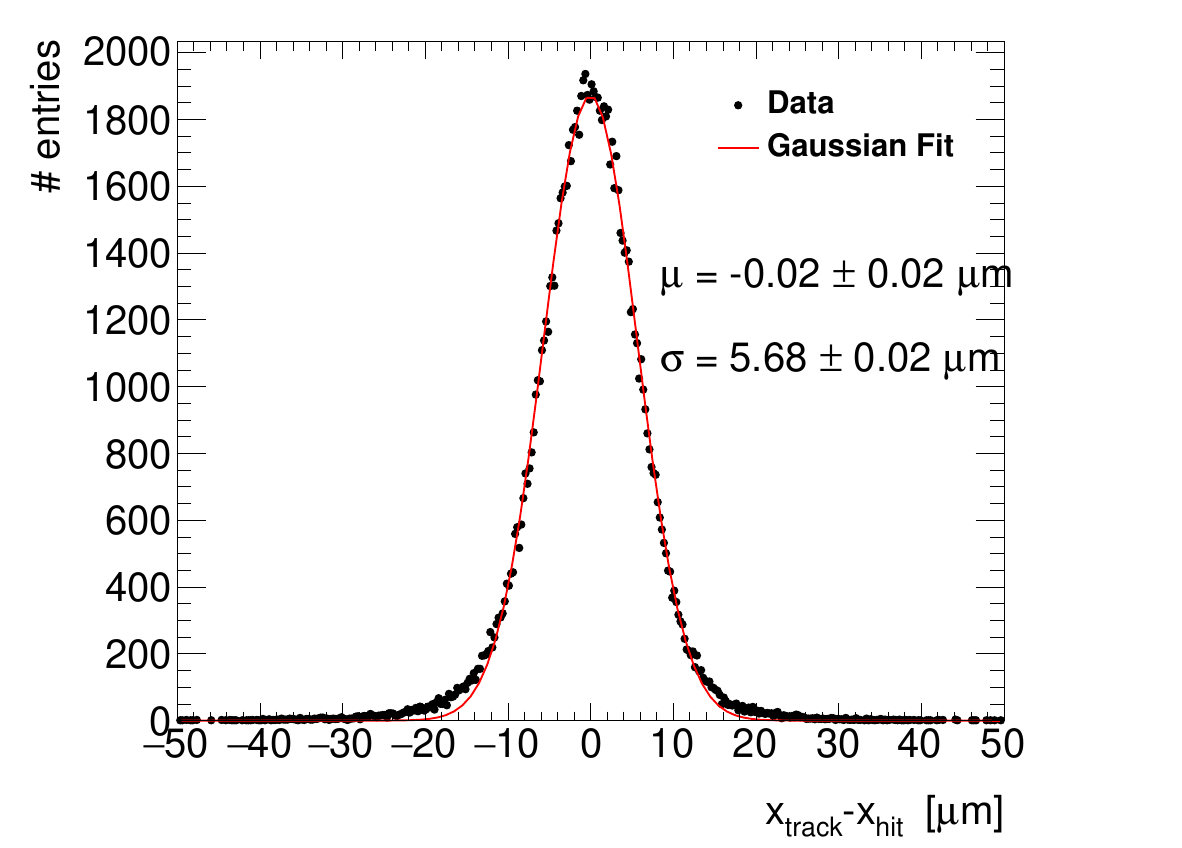} 
        \caption{}
        \label{fig:resX_cs3}
    \end{subfigure}
    \hfill
    \begin{subfigure}[b]{0.495\textwidth}
        \centering
        \includegraphics[width=\textwidth]{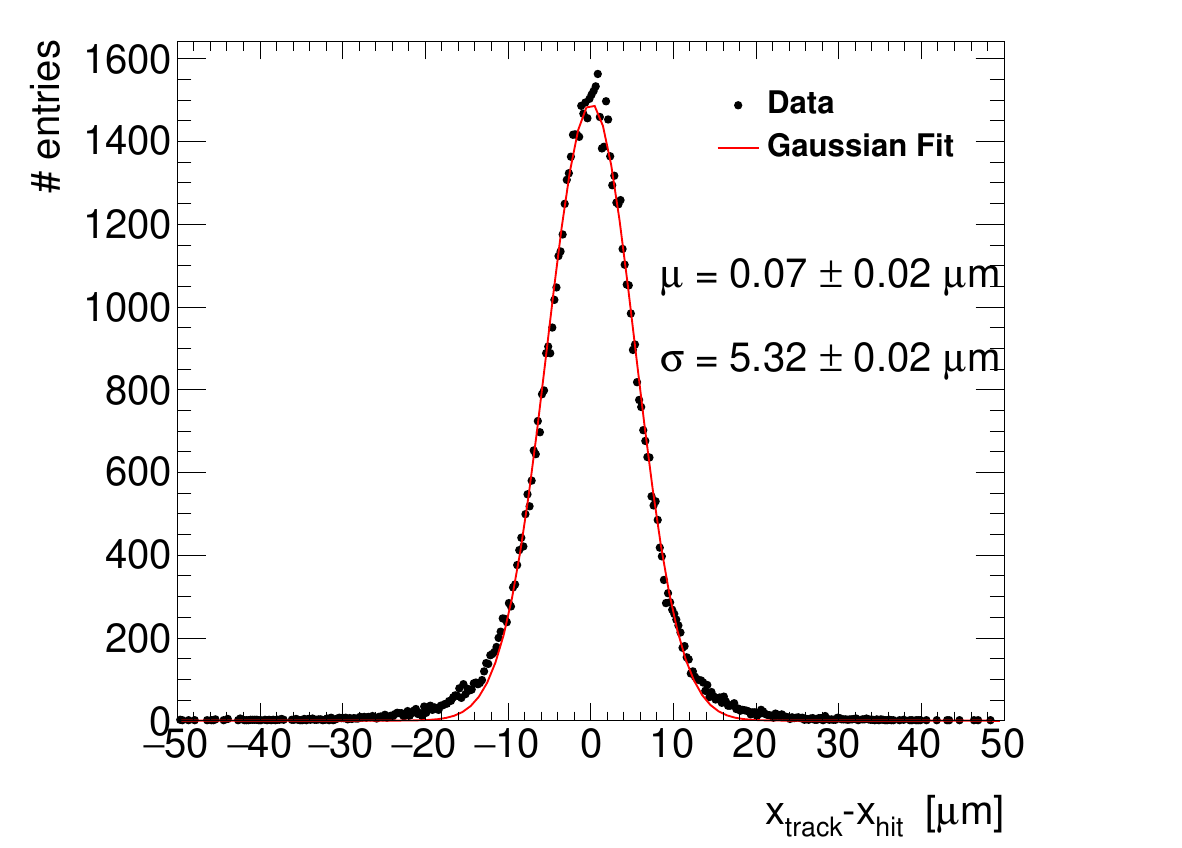} 
        \caption{}
        \label{fig:resX_cs4}
    \end{subfigure}
    \caption{Distribution of residuals in the \(x\) direction for different cluster sizes: (a) CS = 1. (b) CS = 2. (c) CS = 3. (d) CS = 4. Experiment parameters: beam energy = \SI{5.8}{\GeV}, threshold = \SI{200}{\electron}.}
    \label{fig:resX_cs}
\end{figure}

\begin{figure}[htbp]
    \centering
    \begin{subfigure}[b]{0.495\textwidth}
        \centering
        \includegraphics[width=\textwidth]{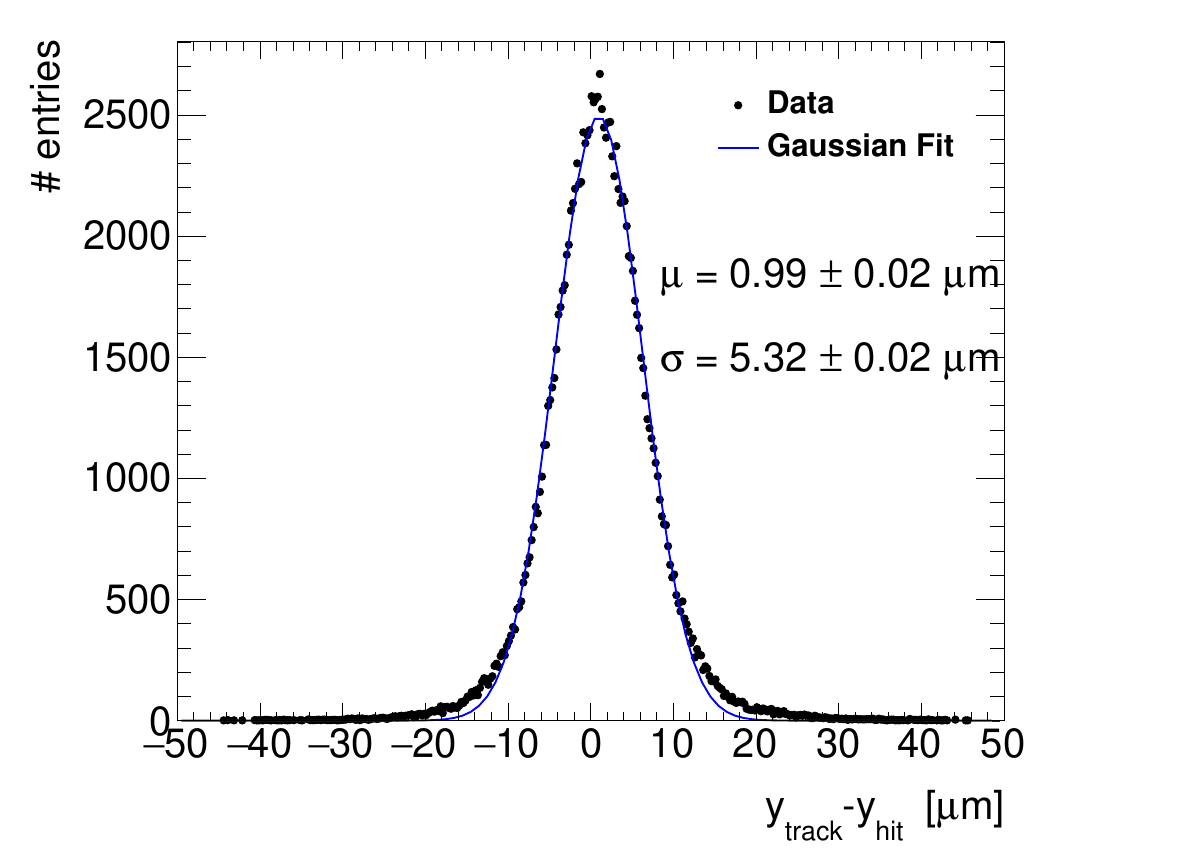} 
        \caption{}
        \label{fig:resY_cs1}
    \end{subfigure}
    \hfill
    \begin{subfigure}[b]{0.495\textwidth}
        \centering
        \includegraphics[width=\textwidth]{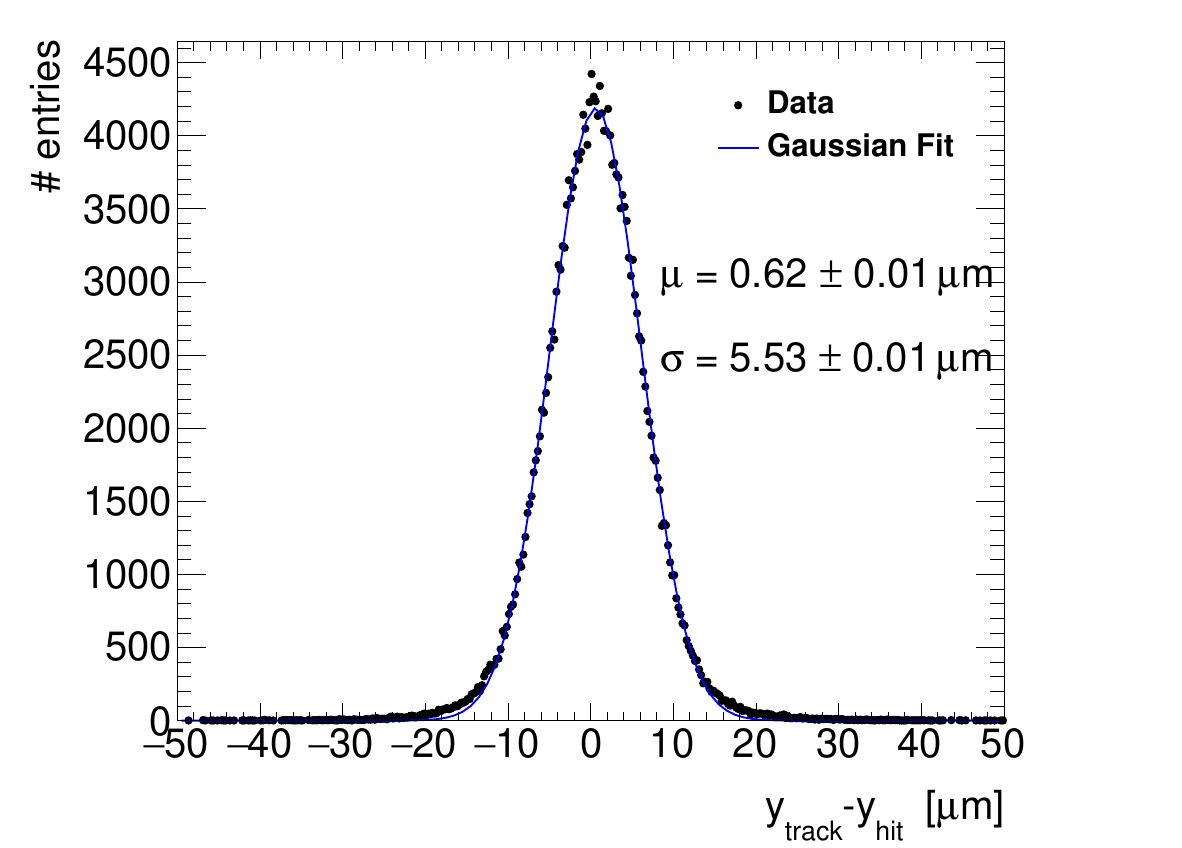} 
        \caption{}
        \label{fig:resY_cs2}
    \end{subfigure}
    \centering
    \begin{subfigure}[b]{0.495\textwidth}
        \centering
        \includegraphics[width=\textwidth]{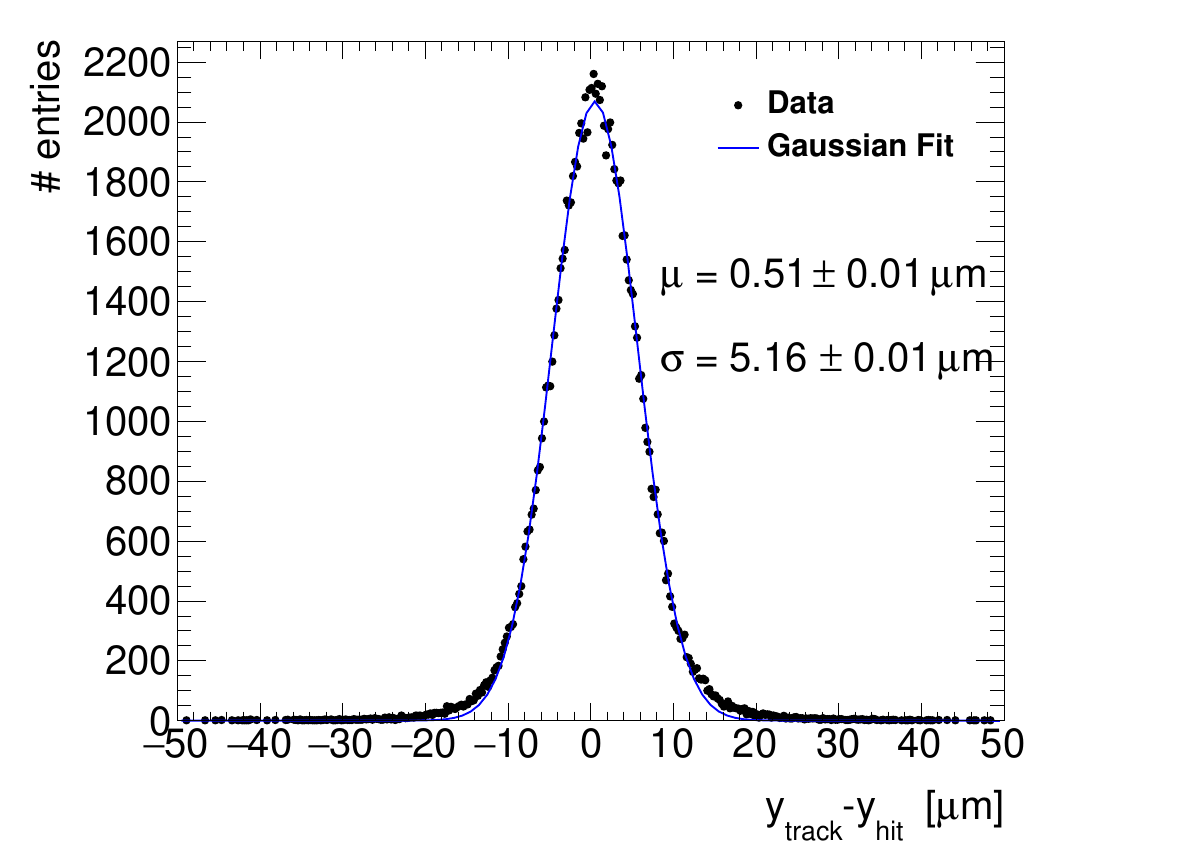} 
        \caption{}
        \label{fig:resY_cs3}
    \end{subfigure}
    \hfill
    \begin{subfigure}[b]{0.495\textwidth}
        \centering
        \includegraphics[width=\textwidth]{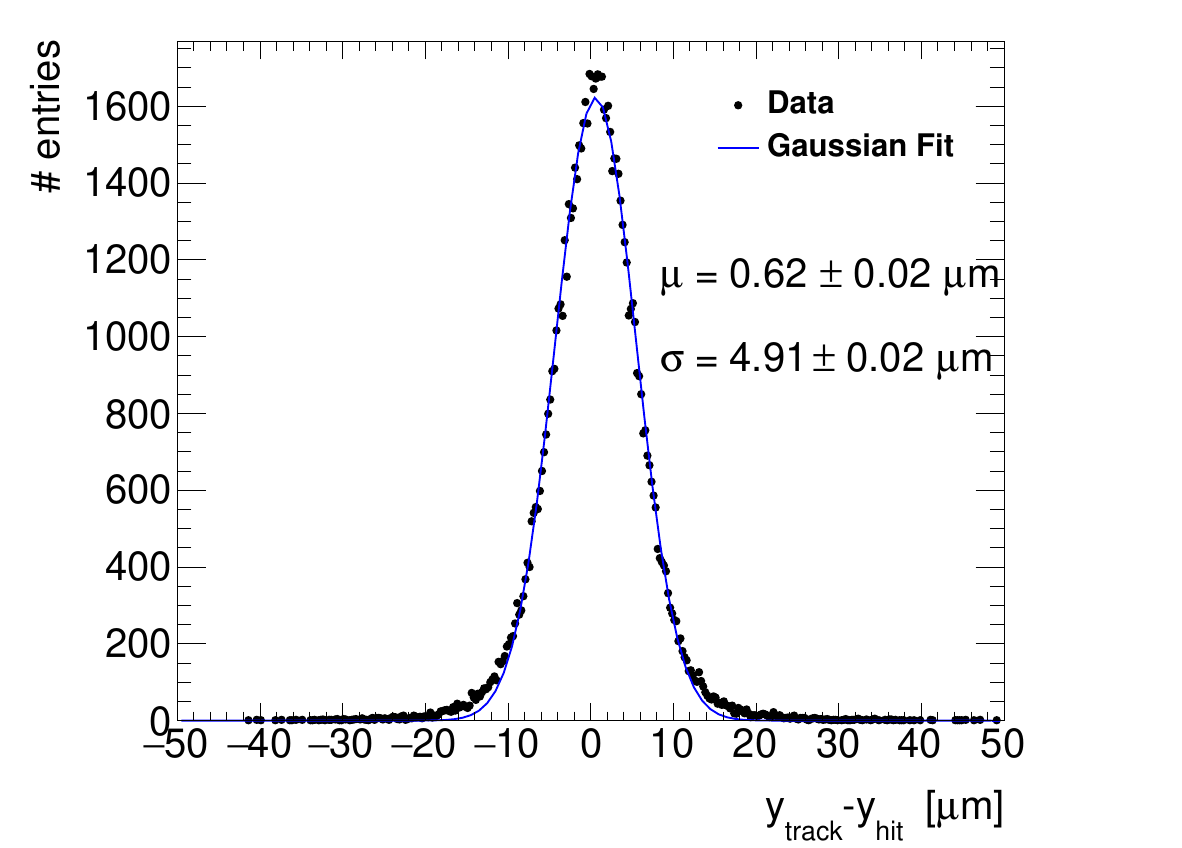} 
        \caption{}
        \label{fig:resY_cs4}
    \end{subfigure}
    \caption{Distribution of residuals in the \(y\) direction for different cluster sizes: (a) CS = 1. (b) CS = 2. (c) CS = 3. (d) CS = 4. Experiment parameters: beam energy = \SI{5.8}{\GeV}, threshold = \SI{200}{\electron}.}
    \label{fig:resY_cs}
\end{figure}

The unbiased residual distributions in Figure~\ref{fig:res_0} have also been plotted for CSs ranging from 1 to 4 on the DUT. The measured spatial resolutions in the \(x\) direction, as depicted in Figure~\ref{fig:resX_cs}, are 9.3, 7.2, 5.7, and \SI{5.3}{\um}, respectively. Similarly, the results for the \(y\) direction are 5.3, 5.5, 5.2, and \SI{4.9}{\um}, respectively, as presented in Figure~\ref{fig:resY_cs}. 

The degraded resolutions can be attributed to the excessive number of tracks outside the favorable regions. As shown in Figure~\ref{fig:cs_map}, the track distributions are always non-zero, which dominates the large residual distribution. 
For instance, with a CS of 1, the tracks are expected to be concentrated in the central region of the pixel. However, a large portion of tracks hit the peripheral region of pixels, as shown in Figure~\ref{fig:cs1_map}, leading to the large residual distribution in Figures~\ref{fig:resX_cs1} and \ref{fig:resY_cs1}.

\FloatBarrier
\subsection{Efficiency and fake-hit rate}
Once all tracks have been reconstructed based on the points from four reference planes, the clusters on the DUT are then associated with the tracks intersecting the DUT. Efficiency is defined as the number of tracks with associated clusters on the DUT over the total number of tracks intersecting the DUT. 

\begin{figure}[h]
    \centering
    \begin{subfigure}[b]{0.495\textwidth}
        \centering
        \includegraphics[width=\textwidth]{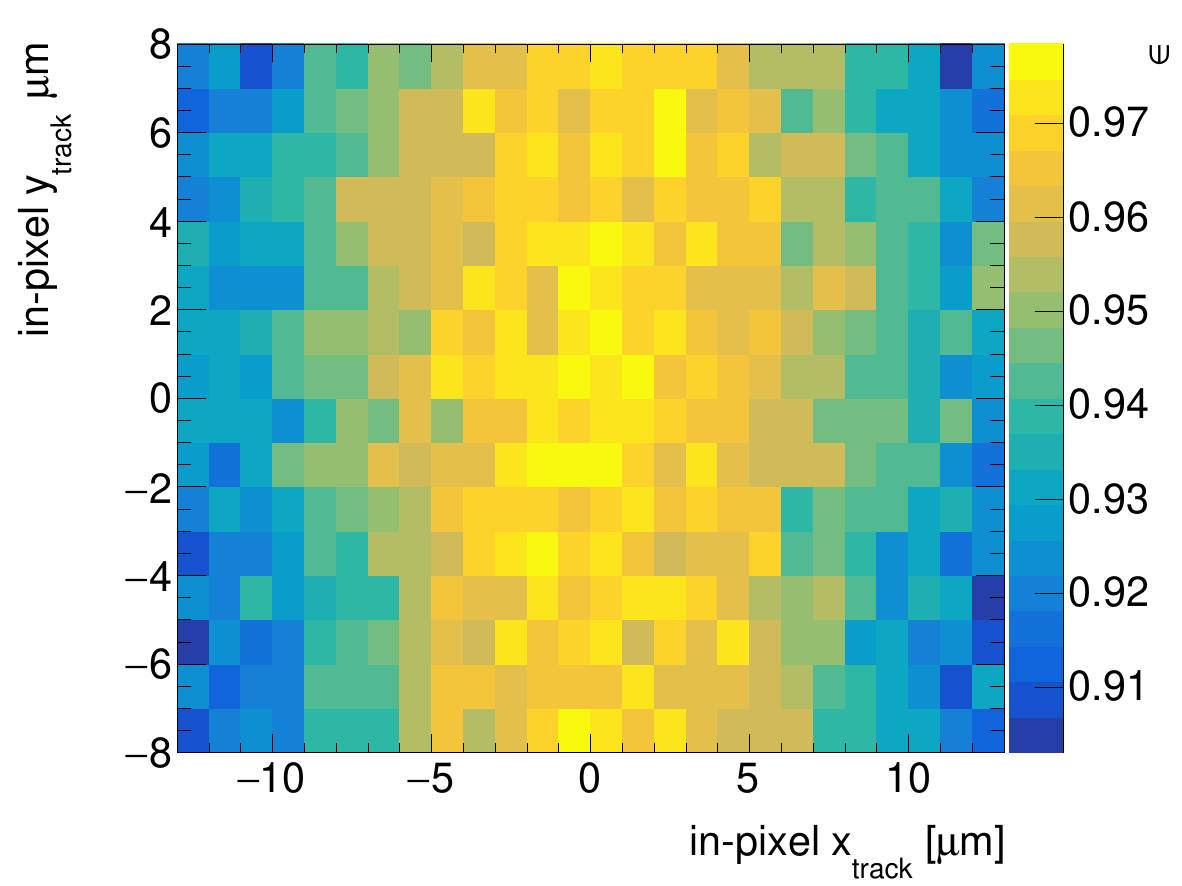} 
        \caption{}
        \label{fig:eff_map}
    \end{subfigure}
    \hfill
    \begin{subfigure}[b]{0.495\textwidth}
        \centering
        \includegraphics[width=\textwidth]{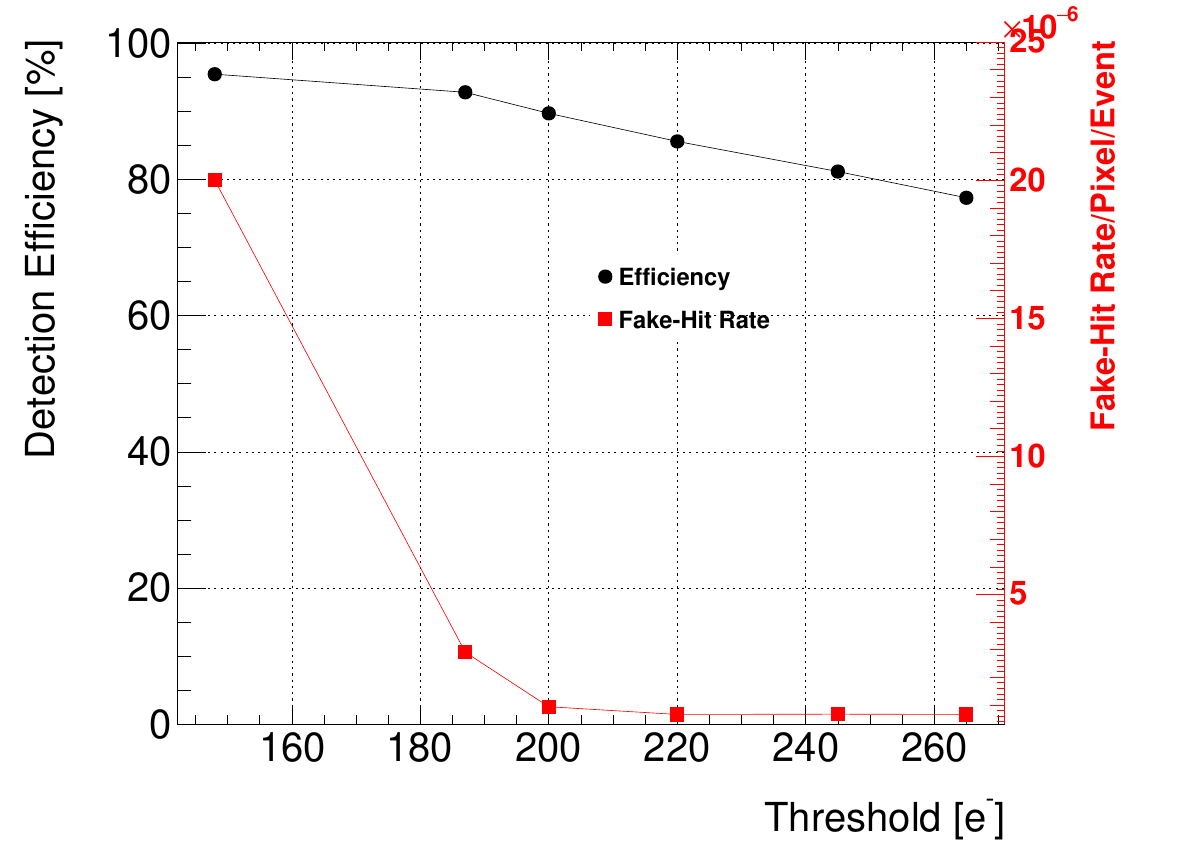} 
        \caption{}
        \label{fig:eff_fhr}
    \end{subfigure}
    \caption{(a) Efficiency plotted as an in-pixel map. Experiment parameters: beam energy = \SI{5.8}{\GeV}, threshold = \SI{200}{\electron}. DUT residual cut = $10*pitch/\sqrt{12}$. (b) The detection efficiency and fake hit rate versus threshold, 26 out of 49k pixels were masked.}
    \label{fig:eff}
\end{figure}

Figure~\ref{fig:eff_map} displays the in-pixel efficiency map, which indicates that the likelihood of a particle being detected by the sensor increases when it hits the center of a pixel. The decrease in efficiency at the short sides is still being investigated, but it may be related to the sharing of charge among more pixels along the short sides of pixels, as shown in Figures~\ref{fig:cs4_map} and \ref{fig:cs3_p2}.

Figure~\ref{fig:eff_fhr} displays the curves of detection efficiency and fake hit rate as the threshold changes. It demonstrates fake-hit rates below $20\times10^{-6}$ per pixel per event, after 26 out of 49k pixels were masked. From the figure, it can be seen that although the detection efficiency decreases with an increase in threshold, the fake hit rate significantly reduces after the threshold exceeds \SI{200}{\electron}. 

\FloatBarrier
\subsection{Threshold scan} \label{sec:threshold_scan}
\begin{figure}[ht]
    \centering
    \begin{subfigure}[b]{0.495\textwidth}
        \centering
        \includegraphics[width=\textwidth]{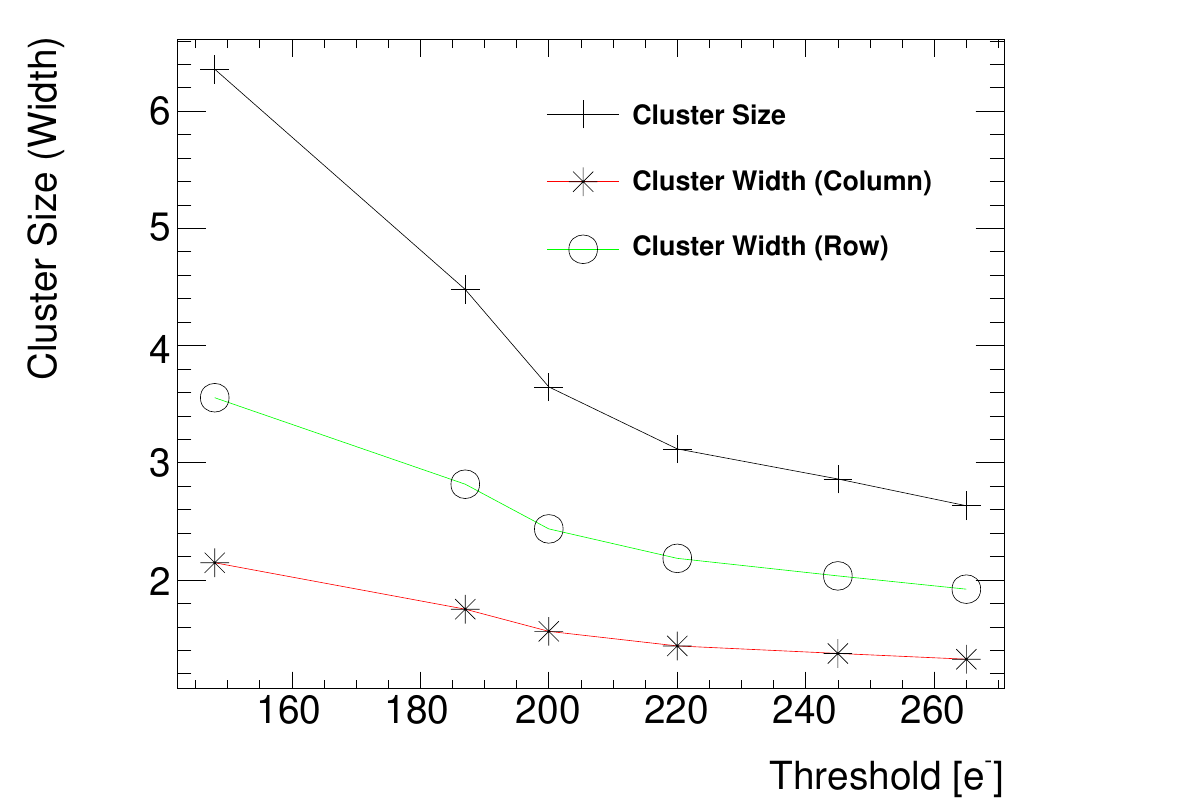} 
        \caption{}
        \label{fig:clusterSizeThresholdScan}
    \end{subfigure}
    \hfill
    \begin{subfigure}[b]{0.495\textwidth}
        \centering
        \includegraphics[width=\textwidth]{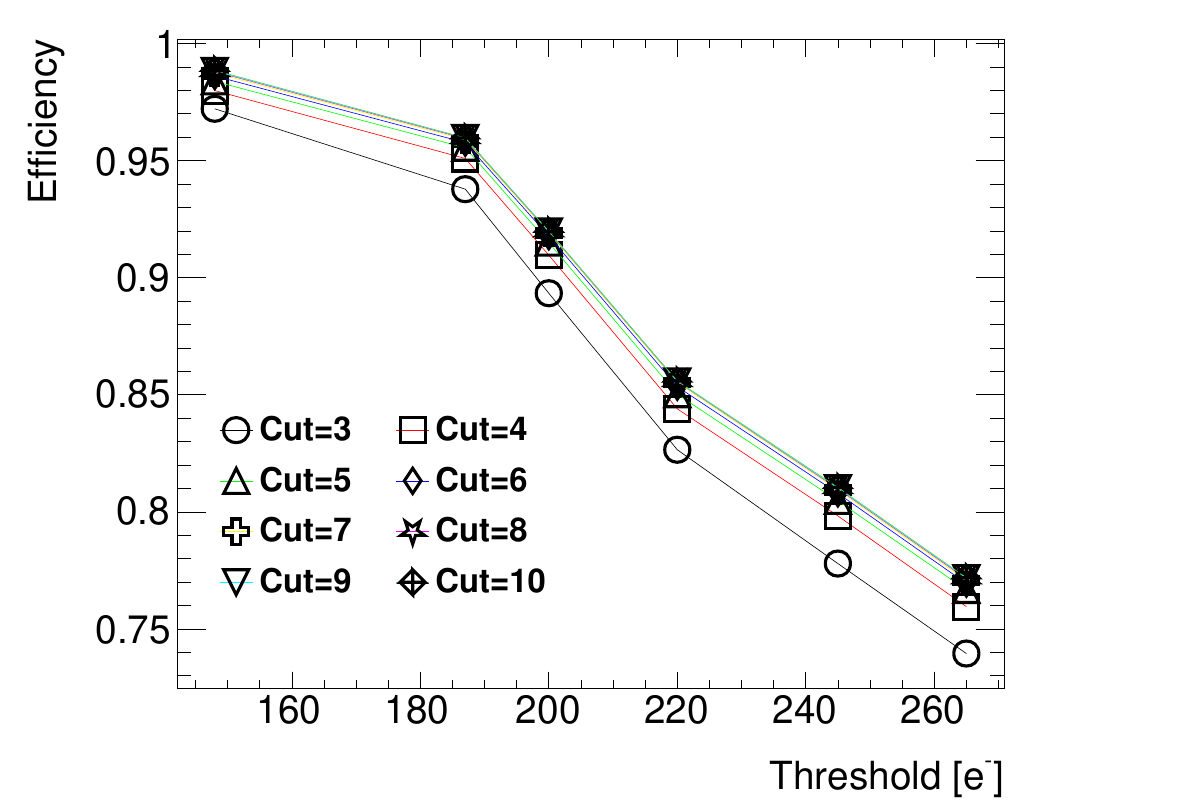} 
        \caption{}
        \label{fig:effThresholdScan}
    \end{subfigure}
    \centering
    \begin{subfigure}[b]{0.495\textwidth}
        \centering
        \includegraphics[width=\textwidth]{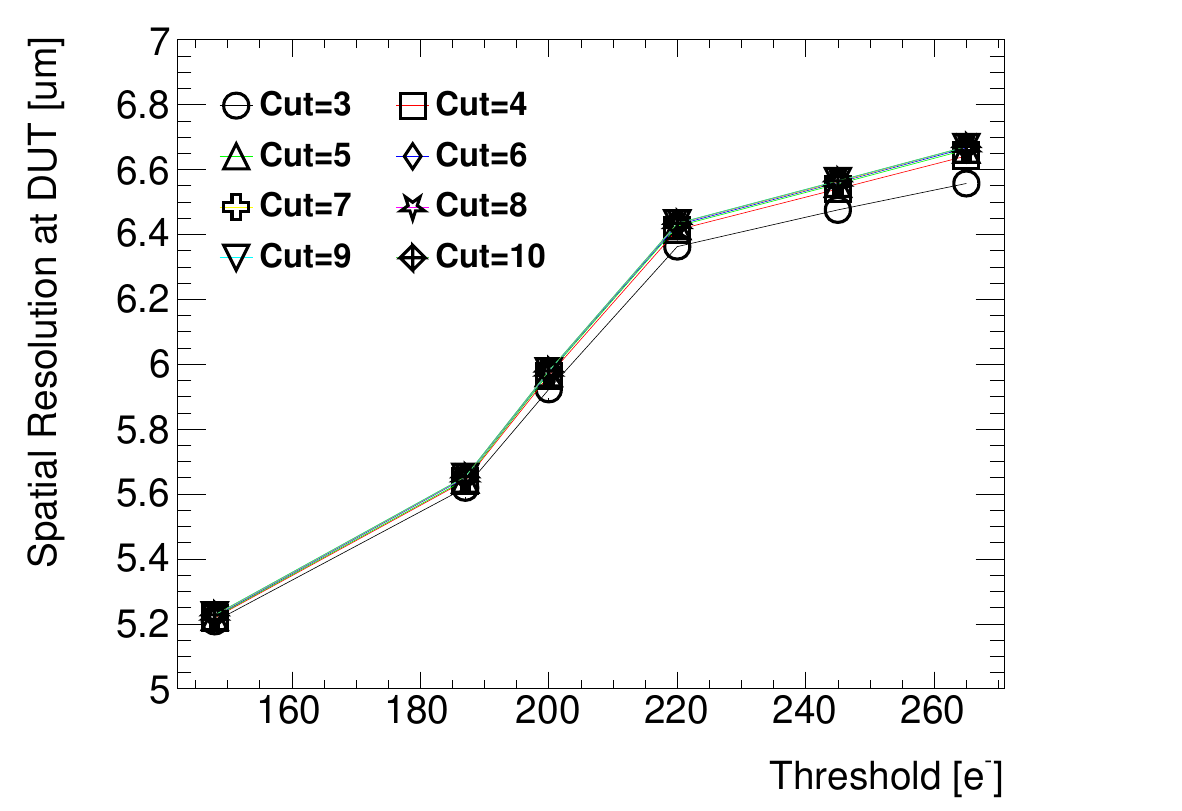} 
        \caption{}
        \label{fig:xResThresholdScan}
    \end{subfigure}
    \hfill
    \begin{subfigure}[b]{0.495\textwidth}
        \centering
        \includegraphics[width=\textwidth]{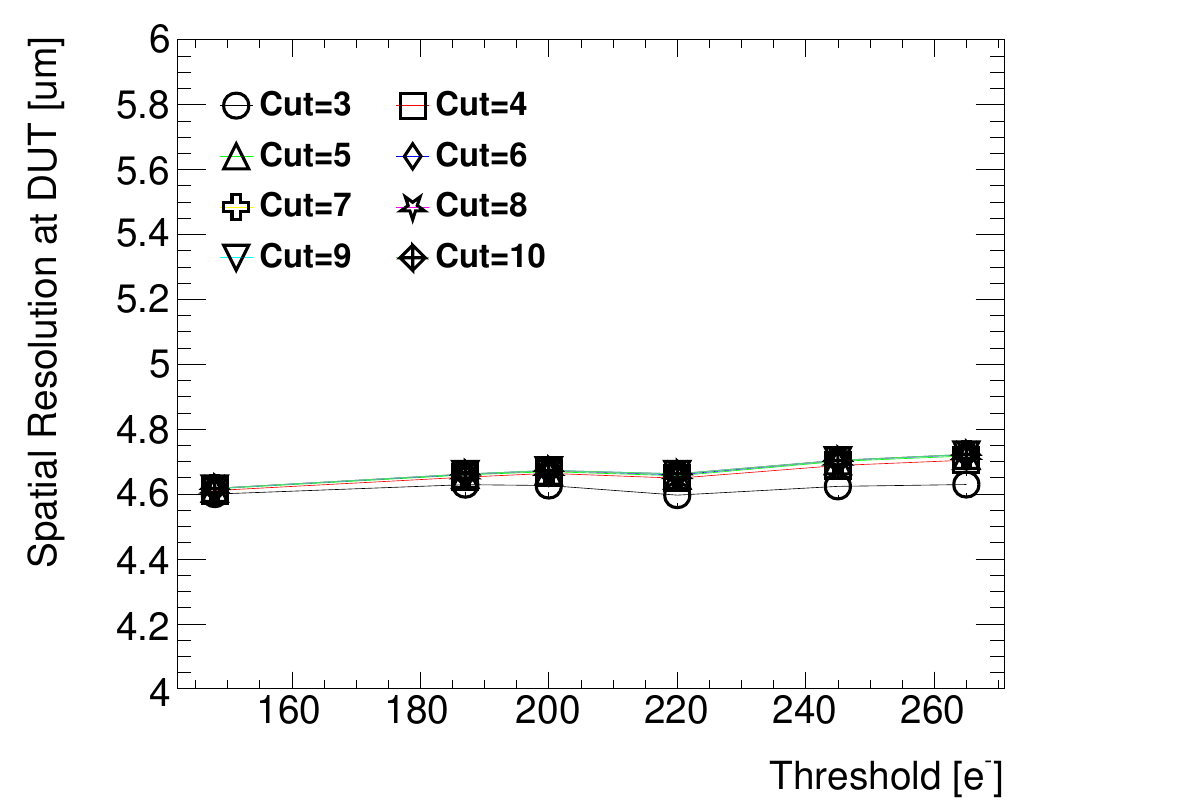} 
        \caption{}
        \label{fig:yResThresholdScan}
    \end{subfigure}
    \caption{Impact of the threshold scan on (a) mean cluster size, (b) efficiency, and (c) resolution in the \(x\) direction and (d) resolution in the \(y\) direction. The DUT residuals are limited to a few times the $pitch/\sqrt{12}$ by applying different cut conditions. Experiment parameters: beam energy = \SI{5.4}{\GeV}, threshold range: from \SI{148}{\electron} to \SI{265}{\electron}.}
    \label{fig:threshold_scan}
\end{figure}

We performed a threshold scan to evaluate changes in several key performance metrics of the sensor, including average CS, detection efficiency, and spatial resolution in both the \(x\) and \(y\) directions. As demonstrated in Figure~\ref{fig:clusterSizeThresholdScan}, an increase in the threshold leads to a gradual decrease in the sensor's average CS, accompanied by a corresponding reduction in detection efficiency, as shown in Figure~\ref{fig:effThresholdScan}. The observed trends are consistent with our expectations. 
For efficiency and spatial resolution, the residual distribution of DUT have been cut according to its quantity. The unit is in the number of times of the $pitch/\sqrt{12}$.
A larger cut size results in higher detection efficiency.

In Figure~\ref{fig:xResThresholdScan}, it is evident that within the threshold scanning range, a higher threshold degrades the spatial resolution in the \(x\) direction. In Figure~\ref{fig:yResThresholdScan}, the degradation of spatial resolution in the \(y\) direction is less significant. 
This phenomenon can be explained by an increase of the polulation in Figure~\ref{fig:resX_cs1} with the average size in the x-direction reducing from 2.2 to 1.3. In contrast, the residual distribution in Figure~\ref{fig:resY_cs} does not change much across cluster sizes 1 to 4, therefore the average cluster size in the y-direction has relatively small influence on the spatial resolution.

Based on the threshold scan conducted within this range, setting the JadePix-3 threshold around \SI{200}{\electron} is recommended. This setting not only improves spatial resolution and detection efficiency, but also maintains a low fake-hit rate even at reduced thresholds.

\FloatBarrier
\section{Conclusion}
This study presents the development, testing, and comprehensive analysis of a beam telescope equipped with the JadePix-3 CMOS pixel sensor. The telescope underwent extensive testing at the DESY TB21 facility using an electron beam with energies ranging from 4 to \SI{6}{\GeV}.

An extensive data analysis was performed, including CS, detection efficiency, and spatial resolution. At an electron energy of \SI{5.4}{\GeV}, the telescope demonstrated excellent spatial resolutions of 2.6 and \SI{2.3}{\um} in two dimensions. The study also included a comprehensive investigation of various CSs, focusing on the spatial resolution of the JadePix-3 sensor in the DUT plane.

Through the analysis of a series of threshold scanning results, we determined the optimal operational threshold for the sensor. This allowed us to attain the most favorable spatial resolution and efficiency under these specific threshold conditions.

\section{Acknowledgments}
This study was supported by the National Key Research and Development Program of China (Grant No. 2016YFA0400400), the Strategic Priority Research Program of the Chinese Academy of Sciences (Grant No. XPB23), and the Fundamental Research Funds for the Central Universities of China (WK2360000014).

\section{Declaration of Generative AI and AI-assisted Technologies in the Writing Process}
During the preparation of this work the authors used ChatGPT/OpenAI to rephrase the English wording and expression in formal writing language. After using this tool/service, the authors reviewed and edited the content as needed and take full responsibility for the content of the publication.






\end{document}